\documentclass[twocolumn]{aastex701}

\usepackage{enumitem}
\usepackage{amsfonts}
\usepackage{amsmath}
\usepackage{amssymb}

\usepackage{cleveref}
\usepackage{multirow}
\usepackage{pgfplots}
\pgfplotsset{compat=1.18}
\usepackage{placeins}
\usepackage{tikz}
\usepackage{upgreek}
\usepackage{booktabs}
\usepackage{physics}
\usepackage{pifont}

\usepackage{cancel}
\usepackage{graphicx}
\usepackage{color}
\usepackage{xcolor}
\usepackage{xparse}

\NewDocumentCommand{\adi}{O{blue} m}{\textcolor{#1}{#2}}

\def\F{frw}
\def\I{inv}

\def\vs{\pmb{s}}
\def\vx{\pmb{x}}

\usepackage[T1]{fontenc}

\newcommand{\be}{\begin{equation}}
\newcommand{\ee}{\end{equation}}
\def\lsim{\mathrel{\rlap{\lower4pt\hbox{\hskip0.5pt$\sim$}}
    \raise1pt\hbox{$<$}}}         
\def\gsim{\mathrel{\rlap{\lower4pt\hbox{\hskip0.5pt$\sim$}}
    \raise1pt\hbox{$>$}}}         

\def\ln{{\rm ln}}

\def\lsim{~\rlap{$<$}{\lower 1.0ex\hbox{$\sim$}}}
\def\bsim{~\rlap{$>$}{\lower 1.0ex\hbox{$\sim$}}}
\def\kms{\ {\rm km\,s^{-1}}}

\def\hmpc{\ {\rm {\it h}^{-1}Mpc}}
\providecommand{\lssim}{\mathrel{\lesssim}}
\providecommand{\gssim}{\mathrel{\gtrsim}}

\providecommand{\hhhmpc}{h^3\,{\rm Mpc}^{-3}}

\def\vx{\mathbf{x}}

\def\apjs{{Astrophys.~ J.~ Suppl.~}}
\def\apjl{{Astrophys.~ J.~ Lett.~}}

\shorttitle{2MRS Bayesian Reconstruction}
\shortauthors{A. Nusser}

\begin{document}

\title{Bayesian Reconstruction of the Local Universe from 2MRS: Testing the Gravitational Flow with Cosmicflows-4}

\author[0000-0002-8272-4779]{Adi Nusser}
\email{adin@technion.ac.il}  

\affiliation{The Technion Department of Physics, The Technion – Israel Institute of Technology, Haifa 3200003, Israel}


\begin{abstract}

We present a Bayesian reconstruction of the local density and velocity fields traced by the 2MASS Redshift Survey (2MRS) and test the inferred gravitational flow against independent Cosmicflows-4 (CF4) galaxy-group peculiar velocities.
The fiducial reconstruction is the maximum-a-posteriori (MAP) solution of a Zel'dovich-approximation forward model, constrained by the 2MRS redshift-space distribution through an unbinned Poisson point-process likelihood.
The model assumes Gaussian initial conditions and includes the 2MRS selection function, the Zone of Avoidance, redshift-space distortions, and a distance-dependent galaxy-bias prescription.
The MAP field is obtained by optimization of the posterior, while Hamiltonian Monte Carlo is used to draw posterior samples and constrained realizations within the same framework.
The reconstructed velocity field agrees well with CF4 in object-by-object, density--velocity-correlation, and shell-by-shell reflex-dipole tests.
These comparisons are made at the CF4 redshift-space positions and do not require smoothing the observed CF4 velocities to the MAP resolution.
We also evolve constrained initial conditions with \textsc{Gadget-4}. The real-space density retains the large-scale Zel'dovich structure while developing additional nonlinear small-scale structure, and the redshift-space distribution develops nonlinear Fingers of God.
The results show that the 2MRS field-level reconstruction captures the large-scale gravitational flow of the nearby Universe and provides initial conditions suitable for constrained simulations.

\end{abstract}


\section{Introduction} 
\label{sec:introduction}

Peculiar velocities of galaxies, their motions relative to the
isotropic Hubble expansion, provide a direct probe of the matter
distribution and gravitational dynamics of the nearby Universe. On
large scales, where density fluctuations are linear or mildly
nonlinear, peculiar velocities are sourced by the gravitational field
generated by the density distribution. A comparison between velocities
predicted from a redshift survey and independently measured galaxy
peculiar velocities is therefore a direct test of gravitational
instability theory \citep{Peeb80,NBDL,Nusser2020,LilowNusser2021,carrick15}.

The aim of this paper is to push this gravitational-instability test
beyond previous large-scale comparisons by using the smallest scales on
which a galaxy redshift survey can be modeled reliably. The effective
scale of the comparison should be set by the data and by the dynamical
model connecting density to velocity, rather than imposed in advance by
smoothing the galaxy distribution. We use the 2MASS Redshift Survey
(2MRS) as the density tracer and Cosmicflows-4 (CF4) as an independent
peculiar-velocity catalogue. Exploiting the small-scale information in
the 2MRS distribution requires more than a linear real-space
density--velocity comparison. The model must describe the discreteness of
the galaxy catalogue, the survey selection function and mask,
redshift-space distortions, and an effective nonlinear galaxy-bias
relation. It must also provide a statistical description of the dynamical
mapping from initial density fluctuations to the evolved density and
velocity fields. This requires exploring the landscape of possible
initial fluctuations on all scales, including modes that are only weakly
constrained by the data, under an assumed prior for their statistical
properties.

The main hypothesis tested in this paper can be stated in a
falsifiable form: if the 2MRS field-level reconstruction, together with
the adopted bias, selection, mask, and dynamical model, captures the
large-scale gravitational flow, then the independent CF4 radial peculiar
velocities should be conditionally unbiased at fixed reconstructed
velocity and should show the same density--velocity and shell-by-shell
reflex-dipole correlations as the reconstructed field. Coherent failures
of these tests, beyond the expected CF4 distance errors and nonlinear
small-scale motions, would challenge the interpretation of the
reconstruction as a successful gravitational-instability prediction
within the adopted $\Lambda$CDM model.

This naturally leads to a field-level Bayesian inference problem
\citep{JascheWandelt2013,KitauraEtAl2012,LavauxJasche2016}. We infer
the initial density field on a three-dimensional grid, assign it a
Gaussian prior, evolve it with a differentiable Zel'dovich-approximation
forward model, and compare the resulting redshift-space galaxy
distribution with the observed 2MRS catalogue using an inhomogeneous
Poisson point-process likelihood. The model includes the 2MRS radial
selection function, the Galactic Zone of Avoidance, redshift-space
distortions, and an effective distance-dependent galaxy-bias
prescription.
The MAP field is obtained by optimizing this posterior, while
Hamiltonian Monte Carlo (HMC) is used to draw posterior samples that can
be used as constrained realizations.

The conditional interpretation is important for the CF4 comparison. Let
$V_{\rm Mean}$ denote the posterior-mean velocity field, i.e. the mean
over all velocity fields that are consistent with the observed galaxy
distribution under the assumed prior and forward model. If the prior and
forward model are adequate, this field should be conditionally unbiased,
\begin{equation}
\label{eq:conditional-unbiasedness-mean}
\left\langle V_{\rm true}\mid V_{\rm Mean}\right\rangle
=V_{\rm Mean} .
\end{equation}
This is the relevant unbiasedness property of posterior-mean estimators.
In machine-learning applications, including autoencoder-based
reconstructions, this is the field learned by minimizing a mean-squared-error
loss, whereas minimizing a mean-absolute-error loss returns the conditional
median field instead. Related conditional-bias tests have been emphasized in
machine-learning and Bayesian-reconstruction contexts
\citep{Punya2023,Lilow2024,MundowNusser2025}. In the present inference,
the MAP field is empirically close to the posterior mean. The CF4 test is
therefore interpreted through the corresponding condition
\begin{equation}
\label{eq:conditional-unbiasedness-map}
\left\langle V_{\rm CF4}\mid V_{\rm MAP}\right\rangle
\simeq V_{\rm MAP} ,
\end{equation}
rather than through the reproduction of each noisy individual velocity.
We will also see below that the posterior median field is close to the
MAP field for the comparisons considered here.

This approach is connected to two established lines of work. The first
is the constrained-realization program for the Local Universe, beginning
with the method of \citet{hr91} and developed in the CLUES
program \citep{GottloeberHoffmanYepes2010,CarlesiEtAl2016}. Closely
related 2MRS work used reconstructed peculiar velocities to generate
constrained initial conditions and reproduce major nearby structures
\citep{l10}. The second is field-level Bayesian reconstruction of
cosmological initial conditions and large-scale structure. BORG
\citep{JascheWandelt2013} and \textsc{KIGEN} \citep{KitauraEtAl2012}
showed how galaxy surveys can be used to infer initial conditions and
evolved density and velocity fields with physical forward models. The
BORG analysis of 2M++ by \citet{LavauxJasche2016} is especially
relevant.
Related BORG-based and constrained-realization studies have also
used posterior initial conditions in subsequent simulations and tests of
the Local Universe, so the present work should be viewed as another
application of this broader program rather than as introducing that
strategy.

Several studies have also reconstructed the nearby density and
velocity fields from 2MRS by complementary methods. \citet{LilowNusser2021}
used Wiener filtering, lognormal constrained realizations, and
Poisson-sampled galaxy catalogues to predict the local velocity field
and compare it with external peculiar-velocity data. Machine-learning
approaches trained on mock surveys have also been developed for 2MRS
\citep{Punya2023,Lilow2024}. Those methods learn a mapping from the
observed galaxy distribution to the target density or velocity field, with
a mean-squared-error loss they are naturally interpreted as posterior-mean
estimators, while a mean-absolute-error loss targets the conditional
median. Recent field-level analyses have emphasized that the likelihood
and tracer model are central modeling choices: some approaches use
Poisson or point-process descriptions of galaxy catalogues, whereas
others use Gaussian likelihoods for
gridded fields \citep{McAlpine2025,Kokron2022,Nguyen2024EFT,Bayer2026HEFT,Akitsu2025Likelihood,Spezzati2025Likelihood}.
Generative machine-learning approaches are a complementary route.
They are especially useful when the training simulations include effects
that are cumbersome to write explicitly into an HMC forward model, such as
nonlinear Fingers of God, complex survey selection, and realistic tracer
physics. The tradeoff is that the learned mapping is tied to the training
set and usually has to be retrained when the survey geometry, tracer
model, or cosmology is changed.

The present work writes an explicit posterior and studies
its MAP solution. In the nearly linear regime the posterior
mean, Wiener field, and MAP field are expected to be close, but they are
distinct Bayesian summaries in general.
The distinguishing feature of the present analysis is the combination of
an explicit posterior, HMC inference, an unbinned 2MRS Poisson
likelihood, and a direct external test with CF4. We validate the method
on MDPL2 halo mocks, apply it to 2MRS, and compare the resulting MAP
velocities with CF4 in three ways: point by point through the
conditional mean, through a density--radial-velocity correlation
statistic, and through a shell-by-shell reflex-dipole or bulk-flow
comparison. We also use an HMC posterior draw to construct refined
constrained initial conditions and evolve them with \textsc{Gadget-4}.
This final step is a consistency check on the constrained initial
conditions rather than the primary validation statistic.

The outline of the rest of the paper is as follows. In
Section~\ref{sec:notation-and-conventions} we introduce the notation and the
idealized dynamical reconstruction problem. In Section~\ref{sec:data} we
describe the 2MRS and CF4 data sets. In Section~\ref{sec:bayesian-reconstruction-model}
we define the prior, likelihood, forward model, posterior, and HMC
implementation. In Section~\ref{sec:mdpl2-halo-distribution-and-velocity-comparison} we validate the
method on an MDPL2 halo mock, and in Section~\ref{sec:results} we
apply it to the 2MRS density field and construct constrained initial
conditions. In Section~\ref{sec:object-by-object-cf4-velocity-comparison} we test
the reconstructed velocity field against CF4, and in Section~\ref{sec:discussion}
we summarize the implications and limitations.

\section{Notation and conventions}
\label{sec:notation-and-conventions}
Our notation is as follows. The scale factor is
$a=(1+z)^{-1}$, where $z$ is the cosmological redshift, and $H(z)$ is
the Hubble expansion rate at redshift $z$. The growth factor of the
linear growing mode is denoted by $D(t)$ \citep{Peeb80} and is
normalized such that $D(t_0)=1$ at the present time. The linear growth
rate is
$f \equiv \frac{d\ln D}{d\ln a}$, and we adopt the approximation
$
f \approx \Omega_m^{0.55}
$
\citep{Lind05}.

The real-space comoving position of an object at cosmological redshift
$z$ is denoted by $\vx$. Assuming that the observed redshift
$z_\textrm{obs}$ differs from $z$ predominantly because of the
object's peculiar motion, we define its redshift-space comoving
coordinate $\vs$ as
\begin{equation}
\label{eq:redshift-map}
\vs = \vx + H_0^{-1}\, v_r\, \hat{\vx} ,
\end{equation}
where $\hat{\vx}=\vx/|\vx|$, $v_r = \mathbf{v}\cdot\hat{\vx}$ is the radial component of the
comoving peculiar velocity, $\mathbf{v} \equiv \dot{\vx} $, and $s=d_c(z_{\rm obs})$, with $d_c(z)$
the comoving distance corresponding to redshift $z$. 

 Throughout this work we adopt a fiducial $\Lambda$CDM cosmology
with Planck parameters. In particular, this fixes the linear matter
power spectrum $P(k)$ entering the prior and the background
quantities used in the forward evolution.

\subsection{The dynamical reconstruction problem} 
\label{sec:reconstruction-problem}

It is useful to separate the dynamical reconstruction problem from the
observational one. Suppose first that the late-time mass distribution is
known perfectly in real space, with no shot noise, selection effects,
mask, galaxy bias, or redshift-space distortions. Even in this idealized
case, reconstructing the velocity field from the evolved nonlinear
density is a  challenging dynamical problem. In the linear regime the relation is
simple: the density contrast is proportional to the negative velocity
divergence, $\delta\,\propto\,-\nabla\!\cdot\!\vb{v}$, and the velocity
field follows by solving this relation under the assumption of potential
flow. In the nonlinear regime there is no comparably unique local
velocity--density relation. 

The nonlinear velocity-reconstruction problem is most
naturally phrased as a reconstruction of the initial conditions. The
idealized task is to find an initially nearly homogeneous density field
whose gravitational evolution reproduces the  late-time mass
density contrast, $\delta_0(\vx)$. Once this initial field is fixed,
the late-time density and velocity fields are generated together by the
adopted equations of motion. The reconstruction therefore depends on the assumed dynamics, on the
restriction to the growing mode, and on the prior assigned to the initial
fluctuations.

As a dynamical boundary-value problem, this program has seen substantial progress: many algorithms have been developed that generate nearly homogeneous initial conditions whose subsequent evolution reproduces the observed particle distribution with good accuracy. Examples include Peebles' least-action method \citep{Peebles1989}, Bernoulli-equation and Lagrangian reconstruction schemes \citep{NusserDekel92}, fast and iterative action/no-action methods \citep{Nusser2000,BEN02,KN16}, and optimal-transport/Monge--Ampere--Kantorovich approaches, from the original cosmological MAK formulation of \citet{Frisch} to recent applications \citep{Nikakhtar2024}. In the linear regime, the idealized reconstruction reduces to
\[
\delta(t,\vx)=\frac{D(t)}{D(t_{0})}\,\delta_0(\vx),
\]
where $D(t)$ is the linear growth factor. Solutions based on approximate dynamics, and even on full $N$-body evolution, can also be constructed, provided that the early-time evolution is restricted to the linear growing mode so as to exclude the decaying mode and avoid pathological behavior as $t\to 0$.

In general, however, solutions to the boundary-value problem are not unique, and the variationally preferred solution need not coincide with the actual dynamical history, particularly in virialized regions where orbit mixing and multistreaming occur.

There are serious challenges in practical applications of these methods. The
observed galaxy distribution is sparse and flux-limited, which restricts the
range of scales that can be robustly probed in a realistic survey. Moreover,
galaxies are biased tracers of the dark matter distribution, and the bias
relation may itself be non-linear, non-local, and stochastic. A further
complication arises from redshift-space distortions, which make the mapping
between real-space distance and redshift-space position non-unique in
high-density regions, and where incoherent Fingers-of-God motions further
obscure the underlying density field.

Challenges posed by incomplete
data can be partially mitigated by casting the reconstruction problem in a Bayesian framework, by
introducing a statistical prior on the underlying density fluctuations.
In large scale structure studies, this approach was first developed in the context of Wiener filtering
and constrained realizations, which were used to infer the late-time density
and velocity fields from observations in the linear regime under the
assumption of Gaussian fluctuations \citep{hr91,zh95,Fisher95b}. These methods were
not aimed at reconstructing the initial conditions, but rather
at recovering the evolved large-scale fields that are most consistent with
the data and with a prior specified by a model power spectrum. In this
framework, an estimate of the underlying field is obtained by maximizing
the posterior probability of the field given the data. For Gaussian priors
and Gaussian likelihoods, the maximum-a-posteriori (MAP) estimate
coincides with the posterior mean field and the Wiener filtered field
\citep{zh95}. The method of \citet{hr91} then goes beyond the
MAP/mean/Wiener field by efficiently sampling the Gaussian posterior,
thereby generating random constrained realizations consistent with the
data.

The Bayesian framework is, however, not confined to the linear regime
or to the recovery of late-time fields. When a forward dynamical model
is available to evolve candidate initial conditions to the observed
epoch, the same probabilistic approach can be extended to infer the
primordial density field beyond the linear regime. Such an extension
also requires a prescription for relating the dark matter distribution
to the observed galaxy distribution, through a bias model and an
appropriate likelihood. Because the resulting posterior is
high-dimensional and non-Gaussian, efficient sampling becomes a
substantial challenge. HMC based methods
\citep{JascheWandelt2013,LavauxJasche2016,KitauraEtAl2012} provide a
natural approach for sampling the posterior landscape.

\section{Data}
\label{sec:data}
\subsection{2MRS}
\label{sec:data-2mrs}
The 2MRS is a flux-limited survey with a $K_s$-band magnitude limit of
$K_s \leq 11.75$, providing sky positions and spectroscopic redshifts
for $44,572$ galaxies \citep{2mrs2012,Macri2019}. Its angular footprint covers
$91\%$ of the sky, with the main missing region being the Galactic
Zone of Avoidance (ZoA).

\subsection{Cosmicflows-4}
\label{sec:data-cf4}
As an external velocity test, we use the grouped Cosmicflows-4 (CF4)
peculiar-velocity catalogue \citep{CF4}. Cosmicflows-4 compiles
distance estimates for galaxies and galaxy groups from multiple
distance indicators, including Tully--Fisher measurements
\citep{Kourkchi2020}, fundamental-plane distances
\citep{Fundamental_Plan87,Saulder_Fundamental}, Type Ia supernovae
\citep{Riess1998,hht07}, surface-brightness fluctuations
\citep{SBr99}, Type II supernovae, Cepheids, and
tip-of-the-red-giant-branch distances. The catalogue distances can be converted into radial
peculiar velocities, with group averaging used to reduce the impact of
small-scale virial motions and to combine overlapping distance
indicators. The CF4 sampling is not uniform, either on the sky or with
depth \citep{CF4}. The 6dFGSv fundamental-plane sample mainly covers the
celestial south, whereas the much larger SDSS fundamental-plane
contribution is restricted to the celestial and Galactic north. The
Tully--Fisher component is more widely distributed, but also favors
northern coverage. Thus the asymmetry is mild in the nearby volume and
becomes much stronger at larger distances. In the grouped CF4 catalogue
used here, objects with $D\leq100\,{\rm Mpc}$ are only modestly weighted
toward the north, at roughly a $1.3$--$1.5$ north/south ratio depending
on the coordinate system. Beyond $D>100\,{\rm Mpc}$ the imbalance is much
stronger, with about four to six times more objects in the northern
hemisphere, reflecting the fact that abundant high-velocity coverage
beyond $\sim16{,}000\,{\rm km\,s^{-1}}$ is mostly confined to the sector
that is both north celestial and north Galactic. In all comparisons below,
CF4 objects are placed at their observed redshift-space positions, rather
than at the noisy distance-indicator distances. This choice reduces
sensitivity to inhomogeneous Malmquist bias, which arises when distance
errors scatter objects preferentially out of high-density regions along
the line of sight \citep{a82,Nusser2017}. These data are not used in the
posterior and provide an independent comparison sample for the MAP velocity
field.

\subsection{Reference distances for labeled structures}
\label{sec:data-reference-distances}

For orientation in the figures that label nearby structures, we adopt
fiducial distances of $D_{\rm Virgo}=16.5\ {\rm Mpc}$ and
$D_{\rm Coma}=99\ {\rm Mpc}$. The Virgo value is consistent with the
ACS Virgo Cluster Survey surface-brightness-fluctuation distances,
which place the M87 and M49 subclusters at
$16.7\pm0.2\ {\rm Mpc}$ and $16.4\pm0.2\ {\rm Mpc}$, respectively
\citep{Mei2007}. For Coma, published distances are more
method-dependent, with  conservative literature estimates span roughly
$\pm 20\ {\rm Mpc}$ around $100\ {\rm Mpc}$
\citep{deGrijsBono2020}, while recent Type Ia supernova measurements
give $D_{\rm Coma}=98.5\pm2.2\ {\rm Mpc}$ \citep{Scolnic2024}.
We therefore use $99\ {\rm Mpc}$ as the fiducial Coma distance.


\section{Bayesian reconstruction model}
\label{sec:bayesian-reconstruction-model}

We describe our implementation of the Bayesian framework for inferring
the initial density field of the local Universe from the observed
galaxy distribution. The following subsections specify the prior, the
likelihood, the forward dynamical model, and the resulting posterior
sampled by HMC.

We are given observed angular coordinates and redshift-space positions
$\{{\bf s}_i\}_{i=1}^{N_g}$ of $N_g$ galaxies in a redshift survey
with known angular mask and radial selection function. In the
application considered here, the survey is the 2MRS. The sampled
variable in the inference is the whitened initial field
$\epsilon({\bf x})$. The linear density field $\delta_{\rm l}$ is a
deterministic transform of $\epsilon$, and the likelihood input is
obtained only after evolving this field with the forward model. Thus
the posterior can be written as
\begin{equation}
P(\epsilon\mid \{{\bf s}_i\})
\propto
P(\{{\bf s}_i\}\mid \epsilon)\,P(\epsilon) .
\end{equation}
Here $P(\epsilon)$ is a unit Gaussian prior, equivalent to a Gaussian
prior on $\delta_{\rm l}$ with covariance set by the linear matter
power spectrum. The prior, likelihood, and forward model are detailed
in the subsections below.

\subsection{Prior for the initial density field}
\label{sec:prior-initial-density-field}

The sampled field is represented on a cubic grid. Rather than
sampling $\delta_{\rm l}$ directly, we introduce a whitened Gaussian
field $\epsilon({\bf x})$ whose components are independent unit-variance
normal variates,
\begin{equation}
\epsilon({\bf x}) \sim \mathcal{N}(0,1).
\end{equation}
The linear density field in Fourier space is then written as
\begin{equation}
\label{eq:eps_to_delta}
\delta_{\rm l}({\bf k}) = \sqrt{P(k)}\,\epsilon({\bf k}),
\end{equation}
where $P(k)$ is the linear matter power spectrum. By construction,
$\delta_{\rm l}$ is therefore a Gaussian random field with isotropic
two-point statistics and covariance fixed by $P(k)$.

\subsection{The likelihood}
\label{sec:likelihood}

We use two closely related likelihood choices. The main 2MRS
reconstruction uses the unbinned Poisson point-process likelihood for
the observed galaxy positions. For the MDPL2 tests we also consider a
Gaussian TSC-grid likelihood, in which the observed and model tracer
fields are compared after TSC deposition on the same mesh. In both
cases the prior and dynamical forward model are the same, with only the
data-comparison term in the posterior changed.

\subsubsection{Poisson point-process likelihood}
\label{sec:poisson-point-process-likelihood}

 We adopt the point-process likelihood because
it is the natural likelihood for the unbinned galaxy catalogue used in
the main reconstruction. In contrast to a likelihood for gridded counts,
the data term evaluates the model rate at the observed galaxy positions
and avoids choosing a cell-count data vector whose small-scale behavior
can be affected by aliasing, empty cells, and over- or under-dispersion
relative to a Poisson model.

For the Poisson likelihood, the observed galaxy positions in redshift
space are assumed to be a realization of an inhomogeneous point
process. For a given sampled field $\epsilon$, the forward model
described below first constructs the corresponding linear density
field $\delta_{\rm l}$, then evolves it with the Zel'dovich
approximation and the adopted bias prescription to obtain the model
galaxy density contrast in redshift space. The resulting rate per unit
volume for observing a galaxy at redshift-space position ${\bf s}$ is
denoted by $\lambda_\epsilon({\bf s})$. In the idealized limit of no
selection effects and no survey mask, this rate would reduce to the
model galaxy number-density field in redshift space. For a realistic
catalog, the intensity also includes the angular mask and radial
selection function. The detailed numerical construction of this rate is
given in Section~\ref{sec:construction-model-poisson-rate}. Here we write the
corresponding expression schematically as
\begin{equation}
\label{eq:schematic-poisson-rate}
\lambda_\epsilon(\vb{s})
\simeq
n_0\,M(\hat{\vb{s}})\,\phi(r)\,
\left[1+\delta_{g,\epsilon}^{\rm ZA}(\vb{s})\right] \; ,
\end{equation}
where $n_0$ sets the overall intensity normalization, $M$ is the
2MRS angular mask, equal to either zero or unity in the present
implementation, $\phi$ is the radial selection function, and
$\delta_{g,\epsilon}^{\rm ZA}(\vb{s})$ is the nonlinear Zel'dovich
model galaxy density contrast in redshift space generated by the
current $\epsilon$. This field should not be confused with the linear
initial density $\delta_{\rm l}$. The selection function is evaluated at the model real-space distance
$r$, rather than directly at the redshift-space radius $|\vb{s}|$. This
choice is important since using the redshift coordinate in the selection
function would couple the radial selection gradient to peculiar
velocities and produce the so-called Kaiser rocket effect \citep{k87}.

There are also limitations associated with the dynamical and
redshift-space mappings. The Zel'dovich approximation permits shell
crossing, but after shell crossing it does not describe the subsequent
virialization of bound structures. Instead particles do not remain confined to non-linear halos and 
continue on ballistic
Zel'dovich trajectories. Moreover,
the map from real space to redshift space need not be one-to-one in
high-density regions. These issues are treated numerically in the
construction of the Poisson rate described in
Section~\ref{sec:construction-model-poisson-rate}.
              
The point-process likelihood is
\begin{equation}
{\cal L}_{\rm P}(\epsilon)=P(\{{\bf s}_i\}\mid \epsilon)
=
\exp[-\Lambda(\epsilon)]\,
\prod_{i=1}^{N_g} \lambda_\epsilon({\bf s}_i),
\end{equation}
where
\begin{equation}
\Lambda(\epsilon) = \int \lambda_\epsilon({\bf s})\,d^3 s
\end{equation}
is the total expected number of observed galaxies. Equivalently,
\begin{equation}
\label{eq:loglike}
\ln {\cal L}_{\rm P}(\epsilon)
=
\sum_{i=1}^{N_g} \ln \lambda_\epsilon({\bf s}_i) - \Lambda(\epsilon) .
\end{equation}

This is a point-process
likelihood, rather than a likelihood for binned cell counts. The
data enter only through the Poisson rate evaluated at the observed
galaxy positions and through its integral over the survey volume.
Thus, the likelihood is written directly for the galaxy point
distribution itself. The Gaussian likelihood described next is
therefore used only as a diagnostic grid-level alternative, not as the
fiducial catalogue likelihood.

\subsubsection{Gaussian grid-based likelihood}
\label{sec:gaussian-grid-based-likelihood}

For the Gaussian grid-based likelihood, the data vector is a gridded number
density rather than the unbinned point set. The observed tracer
positions are first deposited on the same $N^3$ grid on which the
initial conditions and evolved density field are defined, using the
triangular-shaped-cloud (TSC) assignment scheme. We use
TSC rather than cloud-in-cell (CIC) assignment because the TSC kernel is
smoother, with a continuous
first derivative except at cell boundaries. This makes it better suited
to gradient-based optimization than CIC, whose piecewise-linear kernel
has discontinuous derivatives at cell boundaries. Thus the observed
density in cell
$\alpha$ is,
\begin{equation}
n^{\rm obs}_\alpha
=
\frac{1}{V_{\rm cell}}
\sum_{i=1}^{N_g}
W^{\rm TSC}_\alpha({\bf s}_i),
\end{equation}
where $W^{\rm TSC}_\alpha({\bf s}_i)$ is the TSC weight assigned by
tracer $i$ to cell $\alpha$. This estimate is obtained directly from
the observed positions, with no mask or selection function applied.
The model prediction is evaluated on the
same grid. For the grid likelihood, the
model number density entering the residual is
\begin{equation}
n^{\rm mod}_\alpha
=
\bar n\,M_\alpha\,\phi_\alpha\,
\left[1+\delta_\alpha+\delta_0\right] ,
\end{equation}
where $M_\alpha$ is the angular mask, $\phi_\alpha$ is the selection
function, $\delta_\alpha$ is the model density contrast on the MAP
grid, and $\delta_0$ denotes the optional density monopole term. No
positive transform of $1+\delta_\alpha$ is applied in this likelihood.

The Gaussian log-likelihood is, 
\begin{equation}
\label{eq:normal-grid-like}
\ln {\cal L}_{\rm G}
=
-\frac{1}{2}
\sum_\alpha
\left[
\frac{(n^{\rm obs}_\alpha-n^{\rm mod}_\alpha)^2}
{\sigma^2_{n,\alpha}}
+\ln\left(2\pi\sigma^2_{n,\alpha}\right)
\right] .
\end{equation}
The local rms scatter is assigned from the smooth expected number
density rather than from the fluctuating model density. With the
angular mask handled by restricting the likelihood sum to active cells,
the diagonal TSC shot-noise approximation gives
\begin{equation}
\label{eq:normal-grid-sigma}
\sigma^2_{n,\alpha}
=
c_{\rm TSC}\,\frac{\bar n\,\phi_\alpha}{V_{\rm cell}},
\qquad
c_{\rm TSC}=\left(\frac{11}{20}\right)^3 .
\end{equation}
Thus the Gaussian weight is set only by the mean expected number
density and the radial selection function, and not by the local model
fluctuation $\delta_\alpha$. This expression is a diagonal
approximation to the TSC shot-noise covariance. The full covariance of a TSC-deposited point
process has off-diagonal terms because a single tracer contributes to
several neighboring cells. We neglect these inter-cell covariances and
treat the grid-cell residuals as independent, with the local variance
given by Eq.~(\ref{eq:normal-grid-sigma}).

Thus, rather than using the exact unbinned point-process description,
the Gaussian likelihood defines an approximate grid-level density
likelihood with local shot-noise weighting.

\subsection{Posterior}
\label{sec:posterior}

Collecting all terms, the posterior can be written as
\begin{equation}
\label{eq:posterior}
\ln P(\epsilon\mid \{{\bf s}_i\})
=
\ln{\cal L}_{X}(\epsilon)
-\frac{1}{2}\sum_{\bf x}\epsilon^2({\bf x})\; ,
\end{equation}
where $X={\rm P}$ for the Poisson point-process likelihood and
$X={\rm G}$ for the Gaussian TSC-grid likelihood. For the Poisson
runs, this gives
\begin{equation}
\ln P_{\rm P}(\epsilon\mid \{{\bf s}_i\})
=
\sum_{i=1}^{N_g} \ln \lambda_\epsilon({\bf s}_i)
-\Lambda(\epsilon)
-\frac{1}{2}\sum_{\bf x}\epsilon^2({\bf x})\; .
\end{equation}
The full posterior therefore encodes the tension between the Gaussian
initial-field prior and the observed galaxy data, linked through the
complete forward model.

\subsection{Forward dynamical model}
\label{sec:forward-dynamical-model}

Given $\delta_{\rm l}(\epsilon)$, the forward model constructs the expected
intensity field of galaxies in redshift space. Its central output is
the nonlinear matter distribution generated from the initial field,
which provides the common dynamical input for both likelihoods
considered above. For the Poisson point-process likelihood it is
converted, after redshift-space mapping, selection, mask, and biasing,
into the intensity field $\lambda_\epsilon({\bf s})$. For the Gaussian
grid likelihood it is deposited on the mesh and compared with the
gridded observed density field. The steps are: (i) evolve the initial
field to a nonlinear matter distribution using the Zel'dovich
approximation, (ii) map particles to redshift space when
redshift-space distortions are included, (iii) apply the survey
selection function, (iv) incorporate nonlinear, radius-dependent galaxy
bias, and (v) deposit the weighted particles on a grid to obtain the
model density or Poisson rate required by the chosen likelihood.

\subsubsection{Frame of reference}
\label{sec:frame-of-reference}
The dynamics is modeled with a system of dark-matter particles,
initially placed on a uniform comoving grid inside a periodic box.
Their final positions are obtained using the Zel'dovich approximation,
as described below. Before doing so, however, we must specify the
frame of reference in which particle positions and velocities are
defined. This issue is particularly important because the galaxy
distribution is given in redshift space.

Our local neighborhood, the Local Sheet, within a few
megaparsecs, moves with an almost coherent velocity relative to
the CMB,
\begin{equation}
\mathbf{V}_\textrm{LS}=(381~\kms,\,-331~\kms,\,380~\kms)
\label{eq:local-sheet-cmb-velocity}
\end{equation}
in Supergalactic Cartesian coordinates \citep{Tully2008}. If nearby galaxies were assigned line-of-sight positions in
redshift space from redshifts measured in the CMB frame, this
coherent motion would place them on the wrong side of the
observer relative to their true locations. By contrast, when
redshifts are expressed in the Local Sheet frame, the redshifts of
nearby galaxies provide a much better approximation to their true
distances, thereby avoiding the local anomaly that would
otherwise appear in the redshift-space distribution constructed
from CMB-frame redshifts. Therefore, we work in the reference
frame comoving with the Local Sheet. In this frame, matter at
larger distances should exhibit a reflex bulk flow associated with
the motion of the Local Sheet relative to the CMB.

\subsubsection{Zel'dovich evolution}
\label{sec:zeldovich-evolution}

The evolved matter distribution is obtained from the initial density
field using the Zel'dovich approximation \citep{zeld70}. Starting from Lagrangian
grid positions ${\bf q}$, we compute the displacement field
${\boldsymbol \psi}({\bf q})$ from $\delta_{\rm l}$, corresponding to a growth factor $D=1$. In Fourier space,
\begin{equation}
\label{eq:zdisp}
{\boldsymbol \psi}({\bf k}) = i\,\frac{{\bf k}}{k^2}\,\delta_{\rm l}({\bf k}),
\end{equation}
optionally multiplied by a smoothing or filter function. After inverse
Fourier transformation, this yields ${\boldsymbol \psi}({\bf q})$ on the
Lagrangian grid.

The same forward model can also be run with second-order Lagrangian
perturbation theory (2LPT, \citealt{Buchert1993}) by adding the
second-order displacement term. We tested this option for the present
2MRS reconstruction and found only small changes in the recovered
large-scale density and velocity fields relative to the Zel'dovich case.
We therefore use the first-order Zel'dovich model as the default because
it is simpler and less expensive while giving essentially the same
results for the diagnostics emphasized here. This is a deliberately
approximate gravity solver. The present use of ZA is
motivated by the velocity-focused 2MRS--CF4 test and by the limited
resolution at which the nearby redshift survey can be modeled reliably.

The frame choice of \S\ref{sec:frame-of-reference} is implemented
by assigning the Lagrangian grid origin, ${\bf q}=0$, to the Local
Group/Local Sheet frame. We then subtract the Zel'dovich displacement
and velocity evaluated at this grid origin. This is a choice of origin
and inertial frame for the reconstruction. It expresses positions and
peculiar velocities relative to the Local Sheet frame used for the
observed redshifts, rather than identifying a special Eulerian grid node
at $z=0$ or following a distinguished mass element.
Particle
positions are assigned according to
\begin{equation}
\label{eq:zeld}
{\bf x} = {\bf q} + D\,{\boldsymbol \psi}_{\rm rel}({\bf q}),
\end{equation}
where the relative displacement field is
\begin{equation}
{\boldsymbol \psi}_{\rm rel}({\bf q})
=
{\boldsymbol \psi}({\bf q})-{\boldsymbol \psi}({\bf 0}) .
\end{equation}

When redshift-space distortions are included, particle peculiar
velocities are computed from the same displacement field. In the
Zel'dovich approximation, the velocity is proportional to the
displacement,
\begin{equation}
{\bf v} \propto \dot D\,{\boldsymbol \psi}_{\rm rel},
\end{equation}
with $\dot D = H f D$.

The radial direction is defined with respect to the observer at the
box center,
\begin{equation}
\hat{\bf r} = \frac{{\bf x}}{\sqrt{|{\bf x}|^2+r_{\rm soft}^2}},
\end{equation}
where the softening parameter $r_{\rm soft}$ regularizes the
expression near the origin. The radial peculiar velocity is then
\begin{equation}
v_r = {\bf v}\cdot \hat{\bf r}.
\end{equation}
The redshift-space position is modeled as
\begin{equation}
{\bf s}({\bf x}) = {\bf x} + W_{\rm inner}(r)\,v_r\,\hat{\bf r},
\end{equation}
where $W_{\rm inner}(r)$ is introduced to suppress the
line-of-sight distortion very close to the observer and thereby
regularize the mapping. In practice we use a smooth radial taper,
$W_{\rm inner}(r)=1-\exp[-r^2/(2r_{\rm inner}^2)]$, with
$r_{\rm inner}\simeq1\hmpc$. This makes the redshift-space correction
vanish at the origin and approach unity outside the immediate Local
Sheet neighborhood. Thus the same initial density field determines
both the evolved matter distribution and the mapping from real to
redshift space, with all coordinates defined relative to the motion of
the box center.

\subsubsection{Biasing scheme}
\label{sec:biasing-scheme}

Galaxy bias is incorporated through a nonlinear, radius-dependent
mapping from the evolved matter field to an effective galaxy field.
The simple linear prescription,
\begin{equation}
\delta_g = b\,\delta_m ,
\end{equation}
with constant $b$, is problematic because it can yield
$\delta_g < -1$ when $b>1$. A natural positive-definite alternative
is the power-law relation
\begin{equation}
\label{eq:logb}
1+\delta_g = A\,(1+\delta_m)^{b(r)} ,
\end{equation}
where $A$ is chosen such that $\langle \delta_g \rangle = 0$, and
$b(r)$ is allowed to depend on distance from the box center in order
to account for the fact that galaxies observed at larger distances are,
on average, brighter and therefore more strongly biased
\citep[e.g.,][]{Norberg2001,Zehavi2011,carrick15,LilowNusser2021}.
Following \citet{LilowNusser2021}, who inferred a radial bias by
requiring the rms density fluctuations measured at different distances
to be mutually consistent, we use the following compact approximation
in the MAP and HMC runs,
\begin{equation}
\label{eq:bradial}
b(r)=
\begin{cases}
1.25, & r\leq25,\\
1.25+0.005\,(r-25), & 25<r<200,\\
2.125, & r\geq200,
\end{cases}
\end{equation}
where $r$ is in $h^{-1}{\rm Mpc}$.
This is not intended as a general perturbative bias expansion
\citep{Desjacques2018,Nguyen2024EFT,Bayer2026HEFT}
or as a replacement for BORG-like schemes that split
flux-limited catalogues into observed-magnitude bins with separate bias
parameters \citep{JascheWandelt2013,LavauxJasche2016,McAlpine2025}.
Rather, it is an empirical prescription tailored to 2MRS: the
radial dependence of $b(r)$ partially absorbs the luminosity-dependent
tracer population selected by the 2MRS flux limit, while the selection
function $\phi(r)$ accounts for the changing mean number density. 

Although Eq.~(\ref{eq:logb}) guarantees positivity, it is not commutative
with smoothing. That is, the same relation does not in general hold
between galaxy and dark-matter density fields smoothed on scales
different from the grid scale on which the bias is defined.
Moreover, if the bias relation is imposed on small scales where the
rms value of the density contrast is large, it can lead to
unphysically large values of $\delta_g$ for $b>1$. In MAP optimization
or HMC sampling, such large excursions can hinder convergence.

We therefore implement the above power-law biasing in the following
way. We first define a coarse grid whose spacing corresponds to the
scale on which the bias relation is assumed to hold. This scale should
be sufficiently large for the bias model to be meaningful and for the
rms fluctuations to remain moderate. In the implementation below,
we adopt a grid spacing of approximately $10\ \hmpc$.

Once the dark-matter density field has been computed on this coarse
grid by interpolating the real-space Zel'dovich particle positions, we
denote the resulting matter contrast by
$\delta^\textrm{C}_m({\bf x})$. Evaluating \cref{eq:logb} on this grid gives the coarse-grid galaxy
contrast $\delta^\textrm{C}_g({\bf x})$, with $A$ chosen so that
$\langle \delta^\textrm{C}_g\rangle_\textrm{C}=0$. Here the superscript
C denotes fields defined on the coarse bias grid, and
$\langle\cdots\rangle_\textrm{C}$ is the volume average over that grid.
The bias ratio is then
\begin{equation}
\label{eq:wbias}
W_{\rm bias}({\bf x}) =
\frac{1+\delta^\textrm{C}_g({\bf x})}{1+\delta^\textrm{C}_m({\bf x})} \; .
\end{equation}
This ratio is then interpolated from the coarse grid to the particles,
yielding a particle weight $W_{{\rm bias},i}$.

The use of the ratio in \cref{eq:wbias} can be understood as follows.
Let the dark-matter particle distribution define the real-space matter
density field. A biased galaxy density field may be written as
\begin{equation}
1+\delta_g({\bf x})
=
W_{\rm bias}({\bf x})\,
\left[1+\delta_m({\bf x})\right] .
\end{equation}
Thus \(W_{\rm bias}\) is the local galaxy-to-matter weight. In the
particle representation, each matter particle carries the interpolated
weight \(W_{{\rm bias},i}\) and is then moved to redshift space using
the map \({\bf s}({\bf x})\) in equation~(\ref{eq:redshift-map}). The
corresponding continuum expression is
\begin{equation}
\label{eq:rsd-weighted-density-continuum}
\begin{aligned}
1+\delta_g^s({\bf s})
={}&
\frac{1}{\mathcal N_s}
\int d^3x\,
\left[1+\delta_m({\bf x})\right]\,
W_{\rm bias}({\bf x})\notag\\
&\times
\delta_D^{(3)}
\!\left[
{\bf s}-{\bf s}({\bf x})
\right] .
\end{aligned}
\end{equation}
where \(\delta_D^{(3)}\) is the three-dimensional Dirac delta function,
and \(\mathcal N_s\) is chosen such that
\(\langle \delta_g^s\rangle_s=0\) over the redshift-space domain used
in the likelihood.

The Dirac delta in equation~(\ref{eq:rsd-weighted-density-continuum}) simply states
that the galaxy weight associated with the real-space mass element at
\({\bf x}\) is deposited at its redshift-space position \({\bf s}({\bf x})\).
Discretizing the integral over the Zel'dovich particles gives
\begin{equation}
\label{eq:rsd-weighted-density-particles}
1+\delta_g^s({\bf s})
\simeq
\frac{1}{\mathcal N_s}
\sum_i
W_{{\rm bias},i}\,
\delta_D^{(3)}
\!\left[
{\bf s}-{\bf s}_i
\right] ,
\end{equation}
up to the common particle mass or volume factor, which is absorbed
into the normalization. On a mesh, the Dirac delta is replaced by the
chosen assignment kernel, for example CIC or TSC,
\begin{equation}
\label{eq:rsd-weighted-density-mesh}
1+\delta_g^s({\bf s}_a)
\simeq
\frac{1}{\mathcal N_s}
\sum_i
W_{{\rm bias},i}\,
W_{\rm assign}({\bf s}_a-{\bf s}_i) ,
\end{equation}
where \({\bf s}_a\) is a redshift-space grid point. Thus the particle
weight \(W_{{\rm bias},i}\) is not an additional assumption, but is the
discrete form of the biased density factor appearing in the
Dirac-delta assignment from real space to redshift space.

\subsubsection{Construction of the model Poisson rate}
\label{sec:construction-model-poisson-rate}

Having defined the Poisson rate schematically in
\cref{eq:schematic-poisson-rate}, we now describe its numerical
implementation on a grid. We suppress the explicit $\epsilon$ subscript on
model-dependent quantities. After the particles have been mapped to
redshift space and assigned bias weights $W_{{\rm bias},i}$, each
particle is also weighted by the selection function evaluated at its
model real-space distance $r_i$. This gives the unnormalized selected
intensity field
\begin{equation}
\rho_\lambda({\bf s})
=
\sum_i \phi(r_i)\,W_{{\rm bias},i}\,
W_{\rm TSC}({\bf s}-{\bf s}_i) \; ,
\end{equation}
where $W_{\rm TSC}$ is the TSC assignment kernel. The selection
function is applied before gridding, at the model real-space position
of each particle, rather than after gridding as a factor
$\phi(|{\bf s}|)$ on the redshift-space mesh. This distinction matters
because particles deposited in the same redshift-space cell can
originate from different real-space distances, especially in
multistream regions. Using $\phi(r_i)$ also avoids coupling the
selection function to peculiar-velocity displacements, which is the
Kaiser rocket effect discussed above.

The model Poisson rate is then taken to be
\begin{equation}
\label{eq:model-poisson-rate}
\lambda({\bf s})
=
n_0\,M(\hat{\bf s})\,
\frac{\rho_\lambda({\bf s})}{\bar\rho_g} \; ,
\end{equation}
where $n_0$ is a normalization constant and $M({\bf \hat s})$ is the
survey mask, which in our case depends only on the angular position.
This is the TSC implementation of
\cref{eq:schematic-poisson-rate}: the ratio
$\rho_\lambda({\bf s})/\bar\rho_g$ is the deposited counterpart of
$\phi(r)[1+\delta_{g}^{\rm ZA}({\bf s})]$, with $\phi$ applied to each
particle at its own real-space radius before deposition. The
normalization by $\bar\rho_g$ uses the volume average of the biased
tracer density before applying the selection function,
$\rho_g({\bf s})=\sum_i W_{{\rm bias},i}W_{\rm TSC}({\bf s}-{\bf s}_i)$,
and the mask ensures that the expected number density vanishes outside
the survey region.

There are two natural ways to fix the normalization $n_0$. One
possibility is to require
\begin{equation}
n_0 \int M({\bf \hat r})\phi(r)\,d^3 r = N_0 ,
\end{equation}
where $N_0$ is the observed total number of galaxies. In this case,
the normalization is fixed by the survey geometry and selection
function, and is independent of the sampled density field. The
likelihood therefore remains sensitive both to the spatial
distribution of the galaxies and to the total number predicted by the
model.
Alternatively, one may impose
\begin{equation}
n_0 \int M({\bf \hat r})\phi(r)\,[1+\delta_g({\bf r})]\,d^3 r = N_0 ,
\end{equation}
so that the total expected number is fixed by construction. This
second choice amounts to conditioning on the observed galaxy count,
thereby removing the information carried by the total abundance and
leaving only the relative spatial distribution to constrain the field.
It also complicates the optimization: although not written explicitly
in the equation, $\delta_g$ is the model galaxy contrast generated by
the current sampled field. Thus $n_0$ becomes an implicit function of
the sampled linear density field, or equivalently of the whitened
field $\epsilon({\bf x})$ used in practice. The normalization would
therefore have to be recomputed at each model evaluation, with this
dependence incorporated consistently in the log posterior. In the
analysis presented here, we adopt the first normalization, based on
$\int M\phi(r)\,d^3 r$, as it produced more robust results.

\subsubsection{Selection function and Zone of Avoidance}
\label{sec:selection-function-and-zone-of-avoidance}
The 2MRS is a nearly all-sky, flux-limited redshift survey, with the
main incompleteness arising from the Zone of Avoidance, defined here
by Galactic latitudes $|b|<5^\circ$. Because the survey is flux
limited, the observed galaxy number density is progressively diluted
with increasing distance. This effect must be incorporated in the
likelihood through a radial selection function $\phi(r)$.

In the forward model, each particle is assigned the weight $\phi(r)$
at its real-space distance $r=|{\bf x}|$. Evaluating $\phi$ at the
real-space rather than the redshift-space position avoids the Kaiser
rocket effect \citep{k87}---the spurious density modulation that can
arise in a flux-limited survey when peculiar velocities move galaxies
across the radial selection function---as discussed in
\S\ref{sec:poisson-point-process-likelihood}.
We adopt the selection function estimate described in
\citep{BND12}.

\subsection{HMC package}

We use \textsc{NumPyro}\footnote{\texttt{https://num.pyro.ai/en/stable/}} \citep{Phan2019NumPyro,Bingham2019Pyro} to implement the log
posterior in \cref{eq:posterior}. The MAP solution is obtained by
minimizing the negative log posterior over the whitened initial field
$\epsilon({\bf x})$ with the \textsc{JAXopt} L-BFGS solver, using the
forward model of Section~\ref{sec:bayesian-reconstruction-model}. The CF4
velocities are not used in this optimization, rather they enter only as an
external validation of the velocity field implied by the MAP density
reconstruction. Posterior samples, when used, are generated separately
with \texttt{NUTS}, the adaptive No-U-Turn variant of Hamiltonian
Monte Carlo \citep{HoffmanGelman2014NUTS}.


\section{Reconstruction using the MDPL2 halo distribution}
\label{sec:mdpl2-halo-distribution-and-velocity-comparison}

As a controlled simulation-level test, we apply the  MAP
reconstruction machinery to halo distributions drawn from the MDPL2
simulation. MDPL2, or MultiDark Planck 2, is a dark-matter-only
$N$-body simulation from the MultiDark Planck suite
\citep{Klypin2016,puebla16}. It follows $3840^3$ particles in a periodic cube of
side length $1\,h^{-1}{\rm Gpc}$, with particle mass
$1.51\times10^9\,h^{-1}M_\odot$, using a Planck cosmology
($\Omega_m=0.307$, $\Omega_\Lambda=0.693$, $h=0.6777$, and
$\sigma_8=0.8228$). The tests below use a cubic MDPL2 subvolume of
side length $L=200\hmpc$ and haloes with
$M_{\rm halo}>10^{12}\,h^{-1}M_\odot$.
    
The purpose here is to assess the quality of the MAP reconstruction from
the MDPL2 halo distribution as an estimator of unfiltered halo peculiar
velocities. Alternatively, one could make the same assessment using the
mean of HMC posterior samples. As we shall see in
Section~\ref{sec:data-map-density-slice}, despite the non-Gaussian nature
of the posterior, the mean field is very close to the MAP field. The MAP
is much easier to obtain than a large ensemble of HMC samples, and hence
we use it as a proxy for the mean field given the data. Therefore, for
our posterior construction to pass this test, the MAP reconstruction must
be conditionally unbiased: the mean actual velocity, without filtering,
should equal the MAP velocity in bins of MAP velocity.
We do not construct flux-limited 2MRS-like mocks  from the
simulation, as in more realistic survey-mock analyses
\citep[e.g.,][]{LilowNusser2021}, and we do not use this test for
parameter estimation. Instead, we ask whether the true halo velocities
are unbiased around binned values of the MAP estimate, and whether
the rms scatter around this estimate is small compared with the
uncertainties of directly observed peculiar velocities. 

 We consider both the
Poisson point-process likelihood and the Gaussian grid-based  likelihood
described above. 
%

The MDPL2 comparison uses halos with
$M_{\rm halo}>10^{12},h^{-1}M_\odot$ in a cubic region of side length
$L=300\hmpc$. This  halo sample has
$\bar n=4\times10^{-3}\hhhmpc$, corresponding to a mean separation
$\bar n^{-1/3}=6.30\hmpc$ and to a sphere of volume $\bar n^{-1}$
with radius $\left(3/(4\pi\bar n)\right)^{1/3}=3.91\hmpc$. We also
consider a randomly diluted sample containing five per cent of the
halos, $f_r=0.05$, for which
$\bar n=2\times10^{-4}\hhhmpc$. The corresponding mean separation and
equal-volume sphere radius are $17.10\hmpc$ and $10.61\hmpc$,
respectively.

Using the \textsc{NumPyro} package, we derive the linear MAP field on a
$128^3$ uniform cubic grid defined in the simulation box. The input to the
reconstruction is the halo distribution after applying redshift-space
distortions. For each MAP solution, the corresponding Zel’dovich-evolved
particle distribution and velocity field are assigned to the grid using
TSC. The MAP radial velocity at each MDPL2 halo position is then obtained
by interpolating the gridded velocity field to the halo’s distorted
position.

\begin{figure*}[t]
\centering
\includegraphics[width=\linewidth]{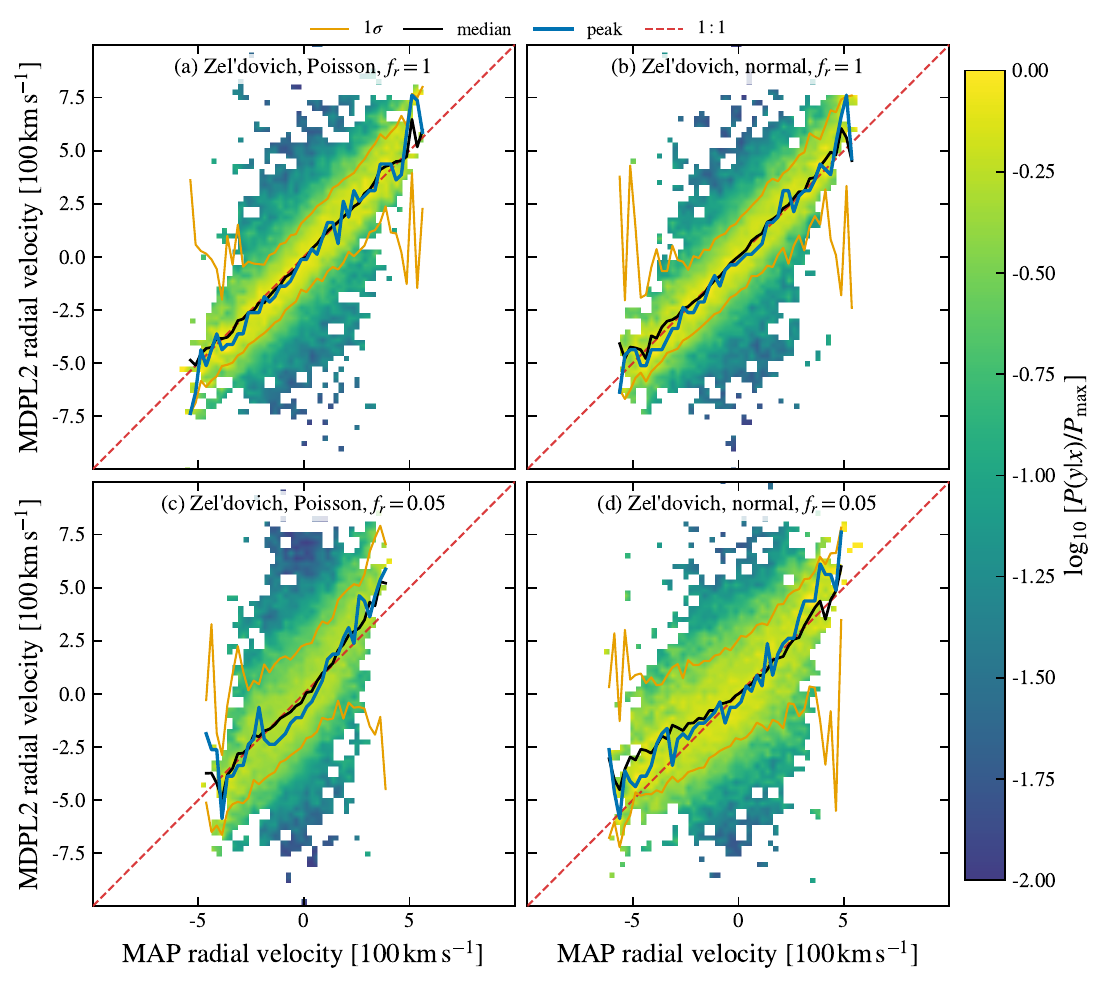}
\caption{MDPL2 conditional heat-map tests of the MAP
$N=128$ radial peculiar-velocity reconstruction. Each panel shows
the conditional distribution of the true MDPL2 halo radial velocity
at fixed MAP radial velocity. The labels Poisson and normal
indicate the likelihood term used in the MAP estimate: an
inhomogeneous Poisson point-process likelihood and a Gaussian
grid-based likelihood, respectively. The top row shows the full
sample, $f_r=1$, while the bottom row shows the diluted sample,
$f_r=0.05$. In each row, the Poisson likelihood is shown on the
left and the Gaussian likelihood on the right.}
\label{fig:mdpl2-zeld-n128-fr1-fr005}
\end{figure*}

\Cref{fig:mdpl2-zeld-n128-fr1-fr005} compares the MAP radial
velocities of MDPL2 halos with their true radial velocities. In each
bin of MAP velocity, the conditional distribution of true velocities is
centered close to the one-to-one relation, shown by the dashed line.
This is seen both from the median relation, shown in black, and from
the blue ridge marking the peak of the conditional distribution. The
agreement is clearest in the top panels, which show MAP
reconstructions from the full sample. Quantitatively, over the
populated velocity bins traced by the overplotted curves, the rms
offset of the median relation from the one-to-one line is $28$ and
$42\kms$ for the full-sample Poisson and Gaussian likelihoods,
respectively, shown in panels (a) and (b). For the diluted sample,
shown in panels (c) and (d), the corresponding offsets are $57$ and
$106\kms$.

We characterize the conditional scatter using the plotted $1\sigma$
envelope. In each populated velocity bin, we measure the half-width of
the central 68 per cent interval about the median and then take the rms
of these half-widths over the populated bins. The resulting values are
$261$, $269$, $317$, and $313\kms$ for panels (a)--(d), respectively.
Thus dilution from $f_r=1$ to $f_r=0.05$ increases the characteristic
conditional scatter from about $260$--$270\kms$ to about
$310$--$320\kms$.

The blue ridge in the heat maps, which marks the conditional peak,
also follows the median closely: the peak--median rms differences are
$71$, $65$, $78$, and $85\kms$ for panels (a)--(d), respectively.
The Poisson and Gaussian likelihoods therefore give similar
full-sample velocity reconstructions. In both the full and diluted
samples, however, the Poisson likelihood gives a smaller median offset
from the one-to-one relation, while the diluted Gaussian case shows the
largest offset.

\section{Application to 2MRS: the density field}
\label{sec:results}

We now apply the method described above to 2MRS, extracting both the
MAP field and samples from the corresponding posterior. In this
application we use only the Poisson likelihood formulation. The
posterior is sampled with HMC in the whitened initial density field,
represented on a $128^3$ Cartesian grid in a periodic box of side
length $L=300\hmpc$, corresponding to a grid spacing of $2.34\hmpc$.
The forward model uses the Zel'dovich approximation with redshift-space
distortions included, and the evolved density is deposited on the same
$128^3$ grid using $128^3$ particles. We use $\Omega_m=0.31$ and growth
normalization $D=1$. The observer is placed at the center of the box,
and only 2MRS galaxies lying within the volume enclosed by the box are
included in the analysis.

The HMC calculation consists of five independent chains, each with
$5000$ warm-up steps followed by $70$ retained samples, giving $350$
stored posterior samples in total. The sampler is run with thinning
factor unity and maximum tree depth $9$. For the density-field summary
figures below, we use the last $30$ retained samples from each chain,
corresponding to $150$ late-chain Zel'dovich-evolved posterior
realizations.

Although the 2MRS catalog extends beyond this volume, the sample
becomes increasingly dilute at larger distances. In particular, fewer
than $5000$ galaxies, out of a total of approximately $43000$, have
redshifts corresponding to $cz/H_0 > 150\hmpc$. We therefore regard the adopted box size and grid resolution as a reasonable 
compromise between volume, resolution, and computational feasibility. Runs on larger grids,  $256^3$, exceeded the available 
GPU memory in the present implementation and, given the small number of 2MRS galaxies beyond $cz/H_0 > 150\hmpc$, would likely offer 
little additional scientific gain.

Because the HMC implementation relies heavily on fast Fourier transforms,
the reconstructed fields are naturally defined with periodic boundary
conditions on the scale of the computational box. This is a technical
choice rather than an essential feature of the reconstruction.
Spherical-Bessel implementations, such as that of \citet{Fisher95b},
allow more flexibility in specifying boundary conditions on a finite
spherical volume, for example by imposing conditions on the density field
at a chosen outer radius. To reduce sensitivity to the periodic boundary
conditions, all comparisons with CF4 are restricted to galaxies within
$120\hmpc$. The remaining buffer between this comparison volume and the
box boundary provides a safety margin that attenuates possible boundary
artifacts.

\subsection{MAP, mean, and median density fields}
\label{sec:data-map-density-slice}

\begin{figure}[!htbp]
    \centering
    \includegraphics[width=\columnwidth]{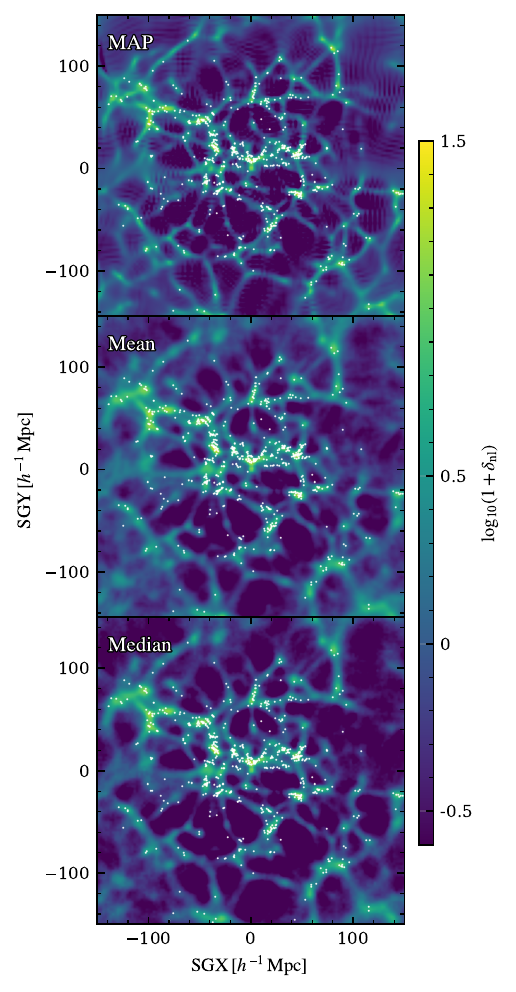}
      \caption{Slice through the Supergalactic plane of the Zel'dovich
redshift-space density field. The panels show, from top to bottom, the
MAP field, the posterior mean of the evolved HMC fields, and the
cell-by-cell posterior median. All panels show
$\log_{10}(1+\delta_{\rm nl})$ with the same color scale. White points
mark 2MRS galaxies selected to lie close to the plane, $|z|/r<0.05$.
Redshifts are in the Local Sheet frame.}
    \label{fig:sliceMAP}
\end{figure}

From the $128^3$ HMC posterior conditioned on the 2MRS redshift-space
catalogue, we compute three summaries of the evolved density field: the
MAP field, the posterior mean, and the cell-by-cell posterior median.
The mean and median are computed from the $150$ selected late-chain HMC
realizations described above.
\Cref{fig:sliceMAP} provides a visual comparison of these summaries in
the Supergalactic plane. The MAP,
posterior mean, and posterior median fields are broadly similar and
trace the same large-scale structures. The posterior mean appears
slightly fuzzier than both the MAP and the median, as expected when
averaging over a finite set of evolved HMC realizations. In the inner
region, roughly within $50\hmpc$, the three fields are very similar and
show comparable small-scale structure. At larger distances, less
small-scale structure is evident because the 2MRS sampling becomes
sparser and the reconstruction is more weakly constrained. The Zone of
Avoidance near $y\simeq0$ is also smoother than neighboring regions,
whereas the observed 2MRS galaxies close to the plane, shown as white
points, trace the main reconstructed structures.

As an additional consistency check, we ask whether posterior
samples from the reconstruction can be used as constrained initial
conditions for a full $N$-body simulation. This test is not used to define
the CF4 velocity comparison. Rather, it checks whether the large-scale
structures constrained by the 2MRS likelihood remain recognizable after
nonlinear evolution and whether the posterior samples contain the
small-scale random power that a single MAP field cannot represent. We
therefore describe below how one HMC sample is converted into a
\textsc{Gadget-4} initial condition.

\subsection{Constrained realization for \textsc{Gadget-4} initial conditions}
\label{sec:constrained-realization-for-gadget4-initial-conditions}

The HMC calculation samples the posterior of the whitened initial
field on a $128^3$ grid from which the linear density $\delta_{128}$ is obtained 
using \cref{eq:eps_to_delta}.

For the \textsc{Gadget-4} run, we refine this HMC realization from
$128^3$ to $256^3$ cells. The goal is to keep the large-scale field
sampled by the HMC chain, while adding only statistically consistent
small-scale power. We therefore require that each coarse cell of
$\delta_{128}$ is equal to the average of the corresponding
$2\times2\times2$ cells in the fine field. In other words, if the fine
field is smoothed back to the $128^3$ grid, it must recover the original
HMC realization:
\begin{equation}
{\cal B}\,\delta_{256}=\delta_{128},
\end{equation}
where ${\cal B}$ denotes block averaging from the $256^3$ grid to the
$128^3$ grid. This constraint is imposed with an algorithm similar to
\citet{hr91}: we first draw a random $256^3$ Gaussian field from the
target linear power spectrum, and then shift its Fourier modes by the
Gaussian conditional mean needed to satisfy the block-average constraint.
The
new modes between the $128^3$ and $256^3$ Nyquist frequencies are
therefore not inferred directly from 2MRS, rather  they are random prior modes
conditioned on the coarse HMC field. The resulting $256^3$ initial
condition preserves the 2MRS-constrained large-scale realization and
adds statistically consistent small-scale structure for the
higher-resolution simulation.

The constrained $\delta_{256}$ field is converted to a normalized
Zel'dovich displacement field using \cref{eq:zdisp}. The initial
conditions for \textsc{Gadget-4} are then obtained by placing particles
on a uniform $256^3$ Lagrangian lattice and displacing them according to
${\bf x}={\bf q}+D_{z}{\boldsymbol \psi}$, where $D_{z}$ is evaluated at
the starting redshift of the simulation, $z=10$. The corresponding
peculiar velocities are assigned using the Zel'dovich relation
${\bf v}=(1+z)^{-1}H_z f_zD_z{\boldsymbol \psi}$ and written, together
with the particle positions and IDs, to a \textsc{Gadget-4} HDF5
initial-condition file.
We start at $z=10$, where the Zel'dovich displacements and
velocities are still small compared with the final nonlinear evolution on
the resolved scales of the $256^3$ run. The mean particle separation is
$dx=1.15\hmpc$, and the adopted softening is well below this scale. 
The initial conditions were evolved to $z=0$ with the \textsc{Gadget-4}
code \citep{Springel2021Gadget4}, using a softening parameter of
$0.2\hmpc$.

One additional bookkeeping step is needed before comparing the final
simulation with the observations and with the MAP reconstruction. The
\textsc{Gadget-4} run is evolved in the fixed frame of the periodic box,
which is the frame most closely associated with the CMB. We therefore do
not force the velocity or displacement at the box center to remain zero
during the run, since that would make the simulation follow the central
mass element instead of staying in the rest frame of the box. The mass
element that corresponds to the Local-Sheet observer can therefore drift
away from the origin during the simulation.

As described in \S\ref{sec:frame-of-reference}, the HMC reconstruction
is written in a Local-Sheet-centered coordinate system. To compare the
two fields fairly, we
shift the whole final simulation box by one periodic translation before
making observer-centered density and velocity maps. The shift is chosen
so that the large-scale $N$-body density field lines up with the
corresponding Zel'dovich density field. This operation does not
change the physics of the simulation. It only changes the coordinate
origin used to display the final snapshot. The density maps below are
shown after this translation has been applied.

\begin{figure*}[!htbp]
    \centering
    \includegraphics[width=.95\textwidth]{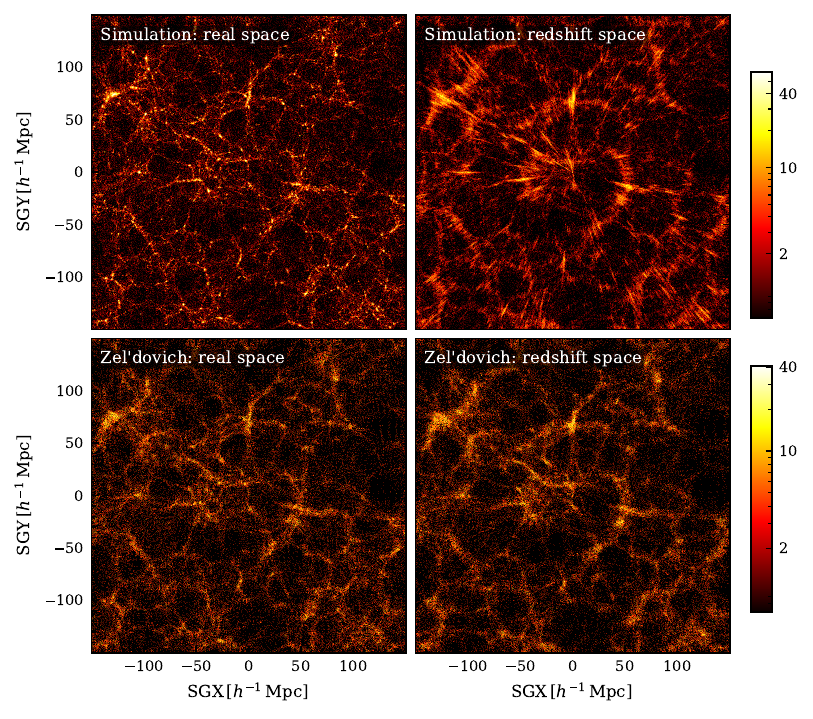}
      \caption{Comparison of projected density maps for one evolved
      $N$-body realization and the corresponding Zel'dovich
      prediction. Each panel shows the ${\rm SGX}$--${\rm SGY}$
      projection through a slab centered on the supergalactic plane
      with $|{\rm SGZ}|<15\,h^{-1}{\rm Mpc}$. Rows show the
      $N$-body and Zel'dovich maps, while  columns show real and redshift
      space. The color scale is the normalized projected density,
      $\Sigma/\langle\Sigma\rangle$. The  pixel size in the
      ${\rm SGX}$--${\rm SGY}$ plane is
      $0.15\,h^{-1}{\rm Mpc}$.}
    \label{fig:single-chain-real-rsd}
\end{figure*}

\begin{figure*}[!htbp]
    \centering
    \includegraphics[width=.8\textwidth]{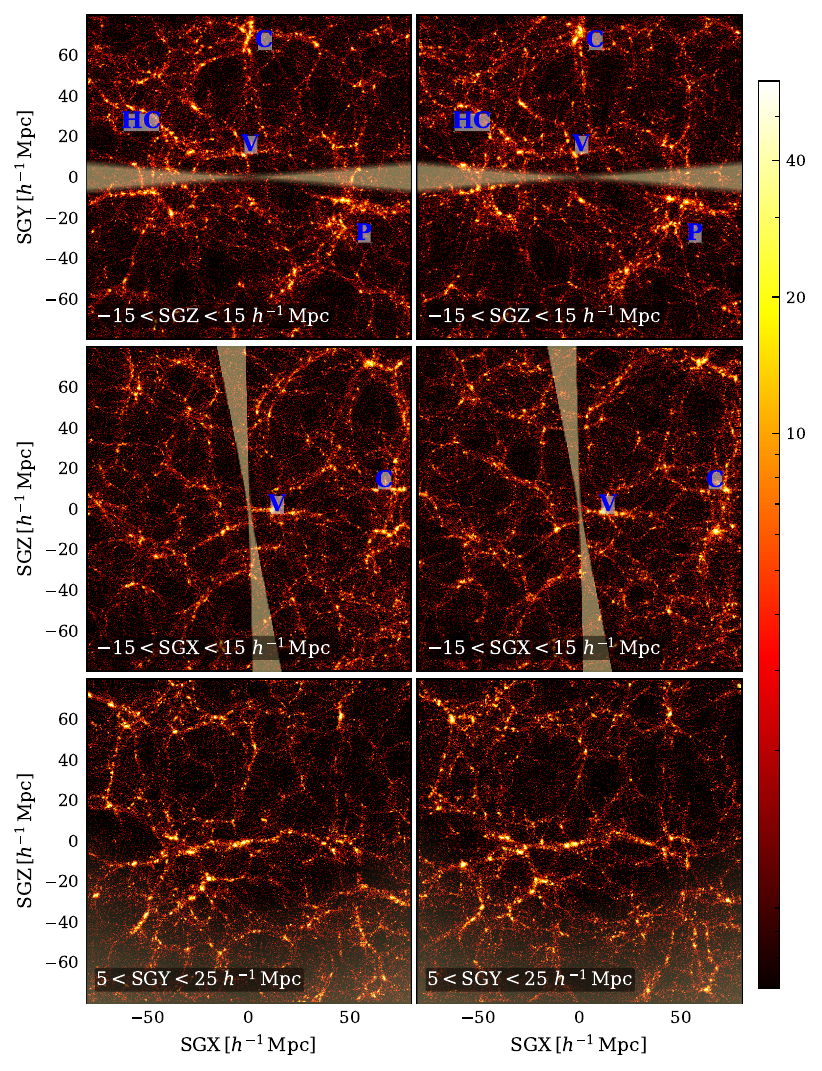}
      \caption{Projected density of two evolved $N$-body realizations initialized from independent HMC posterior draws, in supergalactic coordinates. Columns show the two realizations and rows show the ${\rm SGX}$--${\rm SGY}$ ($|{\rm SGZ}|<15\,h^{-1}{\rm Mpc}$), ${\rm SGY}$--${\rm SGZ}$ ($|{\rm SGX}|<15\,h^{-1}{\rm Mpc}$), and ${\rm SGX}$--${\rm SGZ}$ projections. The ${\rm SGY}$ limits for the bottom row are indicated in the figure. Colors give $\Sigma/\langle\Sigma\rangle$ and yellow shading marks the Galactic zone of avoidance, and blue labels identify selected nearby structures. In the bottom row, fainter yellow shading means that a smaller fraction of the projected slab is obscured by the zone of avoidance. This variation is a geometric effect caused by the inclination of the projected slab relative to the Galactic mask.}
    \label{fig:mask_and_noise}
\end{figure*}

\Cref{fig:single-chain-real-rsd} compares one \textsc{Gadget-4}
$N$-body simulation with the Zel'dovich approximation based on the same
HMC sample. Both maps show the projected density in a slab around the
supergalactic plane. The rows compare the two evolution models. The
columns compare real space with redshift space.

In real space, the two maps agree well on large scales. The main
filaments, dense knots, and voids appear in the same places. The
$N$-body map contains more small-scale structure because it follows
nonlinear gravitational evolution and uses the refined $256^3$ initial
condition. The Zel'dovich map is based on the coarser $128^3$ HMC field.

The difference is clearer in redshift space. The $N$-body map shows
strong Fingers of God that point toward the observer at the center. The
Zel'dovich approximation does not reproduce these virialized features,
because it moves particles along straight-line trajectories. It does,
however, recover the main large-scale redshift-space distortions.

\Cref{fig:mask_and_noise} shows the \textsc{Gadget-4} $N$-body output at
$z=0$ for two simulations with initial conditions constructed from
independent HMC posterior draws, rather than from averages or the MAP
field. These panels show the fully evolved nonlinear particle
distribution in real space and provide examples of nonlinear completions
of the 2MRS-constrained large-scale field. Noticeable differences between
the two simulations are evident in all corresponding panels.

The rows show three projections in supergalactic coordinates. The top row
is the supergalactic plane, ${\rm SGX}$--${\rm SGY}$. The middle row is
the ${\rm SGY}$--${\rm SGZ}$ view. The bottom row is the
${\rm SGX}$--${\rm SGZ}$ view for the ${\rm SGY}$ limits indicated in the
figure, following the supergalactic projection convention used by
\citet{Courtois2013}.  The
labeled structures mark familiar nearby clusters and overdensities,
including Virgo, Coma, Perseus, and Hydra--Centaurus. Their locations are
consistent with the observed nearby Universe at the qualitative level.
This is encouraging because it shows that initial conditions drawn from
the posterior lead to realistic nonlinear structures after
\textsc{Gadget-4} evolution to $z=0$.
The yellow shading marks the Galactic Zone of Avoidance, where the data
constraints are weaker. In the bottom row the yellow overlay is not
uniform. Fainter shading means that a smaller fraction of that part of
the projected slab is hidden by the Zone of Avoidance. This is a
geometric effect caused by the inclination of the slab relative to the
Galactic mask. Compared with the MAP projection, the Zone of
Avoidance in the ${\rm SGX}$--${\rm SGY}$ plane is filled by structures
set by the prior, the dynamical model, and the information propagated
from the surrounding observed regions.

The two posterior draws differ most noticeably on small scales, especially
inside and near the Zone of Avoidance. For example, the filament extending
into the masked region below Virgo in the upper-left panel is much weaker
in the corresponding panel for the second realization. The masked region
above Perseus is also populated differently in the two realizations. Away
from the Zone of Avoidance, the overdense region associated with Coma is
more fragmented in the first realization, whereas the second realization
contains a more concentrated structure. This level of fragmentation is not
unexpected, because the observed large Fingers of God were not collapsed
before applying the reconstruction. The Virgo region shows the opposite
behavior, appearing more concentrated in the first realization than in the
second.

\section{MAP 2MRS vs. CF4 velocity comparison}
\label{sec:object-by-object-cf4-velocity-comparison}

As discussed in Sections~\ref{sec:mdpl2-halo-distribution-and-velocity-comparison}
and~\ref{sec:data-map-density-slice}, the MAP velocity field is a practical
summary of the posterior for the present reconstruction. We therefore test
it against the independent CF4 distance--velocity catalogue by asking
whether the binned mean CF4 velocity follows the MAP prediction at fixed
$V_{\rm MAP}$, rather than whether individual noisy velocities are
reproduced object by object.

\begin{figure}[t]
    \centering
    \includegraphics[width=.92\columnwidth]{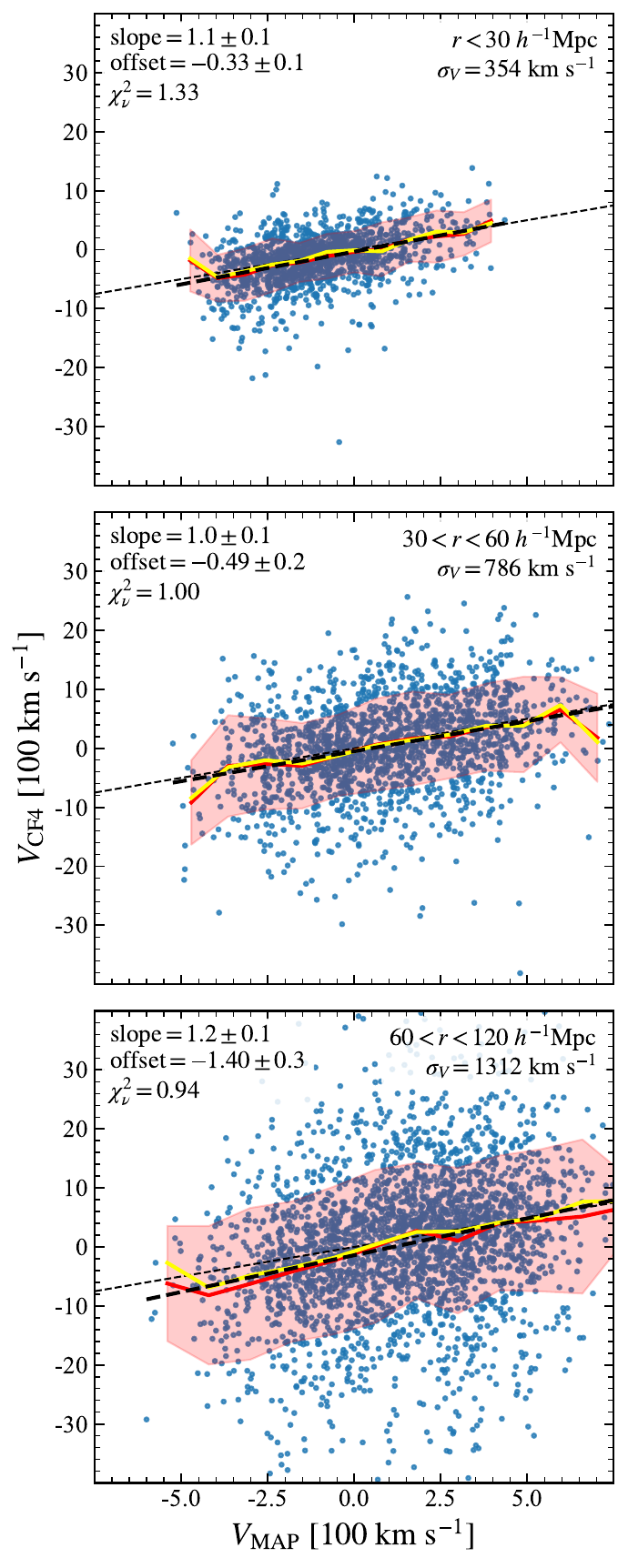}
      \caption{Comparison between observed CF4 radial peculiar velocities, $V_{\rm CF4}$, and 2MRS model-predicted radial velocities, $V_{\rm MAP}$, for the distance ranges indicated in the panels. The red curve and shaded band show the binned mean and scatter. Dashed lines show the one-to-one relation and the best-fitting linear trend.}
    \label{fig:v-v}
\end{figure}

\Cref{fig:v-v} compares the MAP radial velocity prediction with the
independent CF4 peculiar velocities in the distance ranges indicated in
the panels. In each case, the model velocity is obtained by interpolating
the reconstructed field to the CF4 positions and projecting along the
line of sight. The comparison is made by binning in the predicted MAP
velocity. The relevant test is therefore directional. It asks whether the
mean CF4 velocity at fixed reconstructed velocity satisfies
$\langle V_{\rm CF4}\mid V_{\rm MAP}\rangle\simeq V_{\rm MAP}$.
It does not require the individual CF4 velocities to lie close to the
one-to-one line.

The result is strong. In all three distance ranges, the red binned
relation follows the one-to-one line over most of the populated range.
The fitted slopes indicated in the panels are also close to unity. The
offsets of the linear regressions from zero, in units of $100\kms$, are
negligible compared with the rms scatter $\sigma_v$. At the same time,
the individual points form broad vertical clouds. This is expected. CF4
peculiar velocities inherit errors from distance indicators, so their
uncertainties grow roughly in proportion to distance. The scatter
therefore increases from the nearby to the more distant samples.
Unmodeled small-scale coherent motions can add further scatter.

The reduced $\chi^2_\nu$ calculation includes this extra source of CF4
scatter through an rms term $\sigma_0=200\kms$. It is added in quadrature
to the estimated CF4 observational uncertainty. This term is small
compared with the observational uncertainty except in the nearby sample,
where several objects have nearly vanishing formal errors. Raising
$\sigma_0$ to $250\kms$ brings the top-panel value from
$\chi^2_\nu=1.33$ to nearly unity. The effect is negligible in the other
two panels.

The same error model explains why the inverse conditional relation need
not have unit slope. Write the observed CF4 velocity schematically as
\begin{equation}
V_{\rm CF4}=V_{\rm MAP}+\eta ,
\end{equation}
where $\eta$ is a zero-mean random contribution at fixed MAP velocity.
It represents distance-indicator noise and velocity components that are
not captured by the MAP field. The conditional mean can then be unbiased
even when the scatter around it is large. Taking a 
variance $\sigma_\eta^2$ in $\eta$ and variance
$\sigma_{\rm MAP}^2$ in $V_{\rm MAP}$,
\begin{equation}
\langle V_{\rm MAP}\mid V_{\rm CF4}\rangle
\simeq
\frac{\sigma_{\rm MAP}^2}
{\sigma_{\rm MAP}^2+\sigma_\eta^2}\,V_{\rm CF4} ,
\end{equation}
so the inverse slope is attenuated whenever the random contribution is
non-negligible. The important feature of \cref{fig:v-v} is therefore not
the absence of scatter. It is that the binned CF4 mean remains close to
one-to-one despite that scatter.

\subsection{Density--radial-velocity correlation statistic}
\label{sec:density-radial-velocity-correlation-statistic}

\begin{figure}[!htbp]
    \centering
    \includegraphics[width=\columnwidth]{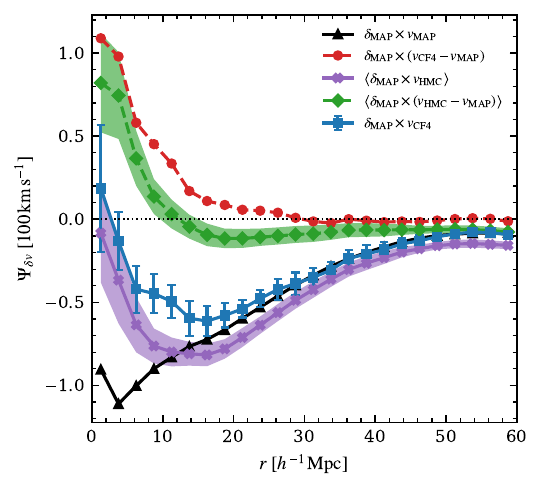}
     \caption{
Density--radial-velocity correlation, $\Psi^s_{\delta v}(r)$, measured
in redshift space with the MAP density field and evaluated for MAP, CF4,
and HMC velocity fields at the CF4 positions. Symbols, curves, and
uncertainty bands are described in the text.
}
    \label{fig:psi}
\end{figure}

The point-by-point comparison is important, but it does not test whether
velocity residuals remain correlated with the reconstructed density field.
If the MAP velocity is a good proxy for the posterior-mean flow, then the
residual velocity should be uncorrelated with the MAP density field under
ideal conditions.

 For a pair consisting of a MAP
density point $i$ and a CF4 object $j$, the positions and separations are
those in redshift space. We write
${\bf x}_{ij}={\bf r}_{2,j}-{\bf r}_{1,i}$,
$x_{ij}=|{\bf x}_{ij}|$, and
$\mu_{ij}=\hat{\bf r}_{2,j}\cdot\hat{\bf x}_{ij}$, where
$\hat{\bf x}_{ij}={\bf x}_{ij}/x_{ij}$. Statistical isotropy and
potential flow imply
\begin{equation}
\left\langle
\delta({\bf r}_{1})\,{\bf v}({\bf r}_{2})
\right\rangle
=
\Psi_{\delta v}(x)\,\hat{\bf x},
\qquad
{\bf x}={\bf r}_{2}-{\bf r}_{1},
\end{equation}
so the radial projection gives
$\langle\delta_i u_j\rangle=\mu_{ij}\Psi_{\delta v}(x_{ij})$, with
$u_j$ denoting the radial velocity used for object $j$. We therefore
estimate the longitudinal amplitude in a separation bin $B_a$ by the
weighted slope
\begin{equation}
\widehat{\Psi}_{\delta v}(x_a)
=
\frac{
\displaystyle\sum_{(i,j)\in B_a}
w_{ij}\,\mu_{ij}\,\delta_i\,u_j
}{
\displaystyle\sum_{(i,j)\in B_a}
w_{ij}\,\mu_{ij}^2
}.
\label{eq:density-radial-velocity-estimator}
\end{equation}
The weights $w_{ij}$ may be used to encode measurement errors or
selection weights, with no additional weighting one sets $w_{ij}=1$.
This normalization deprojects the radial velocities without dividing
individual noisy pairs by $\mu_{ij}$, so nearly transverse pairs carry
little leverage. It is the pairwise least-squares estimate of the
amplitude of the radial velocity component aligned with the separation
vector, and is closely related to the dipole statistic of \citet{Nusser2017}.

\Cref{fig:psi} applies this estimator using the MAP density field for
all curves. The black triangles show the fully reconstructed MAP
correlation, $\delta_{\rm MAP}\times v_{\rm MAP}$, while the blue
squares replace the MAP velocity by the observed CF4 radial velocity,
$\delta_{\rm MAP}\times v_{\rm CF4}$. The red dashed curve shows the
corresponding residual correlation,
$\delta_{\rm MAP}\times (v_{\rm CF4}-v_{\rm MAP})$. Thus the same
density weighting, pair geometry, and estimator normalization are used
for the model and for the data.

The HMC curves provide the posterior comparison for the same statistic.
The purple curve is the mean of
$\delta_{\rm MAP}\times v_{\rm HMC}$ over the HMC velocity
realizations, and the purple band gives the bin-by-bin $1\sigma$
realization-to-realization scatter. The green dashed curve and band show
the same quantity for the residual
$\delta_{\rm MAP}\times (v_{\rm HMC}-v_{\rm MAP})$. These bands
therefore describe posterior/ensemble scatter in the reconstructed HMC
velocities, not the observational CF4 error bars and not the uncertainty
on the mean. To make the comparison with the observed catalogue as
realistic as possible, the HMC/model velocities are evaluated at the CF4
positions and the CF4 radial-velocity error model is propagated through
\cref{eq:density-radial-velocity-estimator}. The blue error bars show
the resulting $1\sigma$ uncertainty on
$\delta_{\rm MAP}\times v_{\rm CF4}$: in each separation bin they are
the standard deviation of Monte Carlo realizations of
$\widehat{\Psi}_{\delta v}$ obtained by adding zero-mean Gaussian
radial-velocity perturbations with the adopted CF4 velocity
uncertainties.

The most direct comparison is between the blue and purple curves, which
correspond respectively to the 2MRS $\delta_{\rm MAP}\times V_{\rm CF4}$
and 2MRS $\delta_{\rm MAP}\times V_{\rm HMC}$ cross-correlations. Both
curves are negative, as expected from mean infall onto dense regions and
outflow from underdense regions. Overall, the agreement is reasonable,
although the CF4 correlation has a lower amplitude, especially at
separations around $10\hmpc$. The correlations approach zero at small
separations, consistent with the stable-clustering regime, which appears
in the Zel'dovich approximation as shell crossing on small scales. At the
largest separations plotted, the cross-correlation is dominated by a
coherent bulk flow. If this bulk component is removed, the two curves
nearly overlap on scales $\gssim 40\hmpc$.

The MAP--MAP cross-correlation, $\delta_{\rm MAP}\times V_{\rm MAP}$
(black curve), matches the CF4 curve well on large scales but deviates
significantly at $r\lssim 10\hmpc$, where small-scale structure and shell
crossing become important. The black curve becomes more negative on these
scales, indicating that the MAP field is dominated by the coherent part of
the flow rather than by virial or shell-crossing motions. The red and
green curves show the corresponding residual correlations,
$\delta_{\rm MAP}\times (V_{\rm CF4}-V_{\rm MAP})$ and
$\delta_{\rm MAP}\times (V_{\rm HMC}-V_{\rm MAP})$. Ideally, both
residual correlations should be consistent with zero, in practice they
are close to zero on scales $\gssim 10\hmpc$.

\subsection{Reflex dipole}
\label{sec:reflex-dipole}
 
For each radial shell of finite thickness, we define the reflex dipole
as the best-fitting constant three-dimensional velocity of the tracers in
that shell, measured relative to the Local Sheet frame and inferred from
their line-of-sight peculiar velocities. 
Explicitly, for tracers in shell $a$ we fit a vector
$\pmb{V}_{\rm ref}(a)$ by minimizing
\begin{equation}
\chi_a^2=\sum_{i\in a} w_i\left[u_i-\pmb{V}_{\rm ref}(a)\cdot
\hat{\pmb{r}}_i\right]^2,
\end{equation}
where $u_i$ is the observed or model radial peculiar velocity and
$w_i$ is the inverse-variance weight when CF4 velocity errors are used
(and unity otherwise).

This quantity is of interest
because, in the ideal limit of
full angular coverage, the dipole component 
is determined by the
dipole moment of the mass distribution 
enclosed by the shell
\citep{ND94}. At sufficiently large radii, 
the reflex dipole is
therefore expected to approach the negative of the CMB dipole
associated with the Local Group motion relative to the CMB.

This comparison uses the HMC realizations rather than only the MAP field
because the reflex dipole is a large-scale, window-dependent statistic.
The finite reconstruction box also removes modes larger than the
modeled volume, so this statistic is not interpreted as a complete
prediction of the asymptotic CMB dipole. It is a finite-volume test using
the same shell window and tracer positions as the data.
The MAP gives a single conditional optimum and is the right tool for the conditional 
velocity--velocity  comparisons above, but it does
not describe the allowed scatter from weakly constrained modes especially at large distance.
 The HMC samples propagate
these sources of posterior uncertainty into the same shell estimator used
for the data, while retaining the MAP as the central reconstruction around
which the constrained realizations fluctuate.

\begin{figure}[t]
    \centering
    \includegraphics[width=\columnwidth]{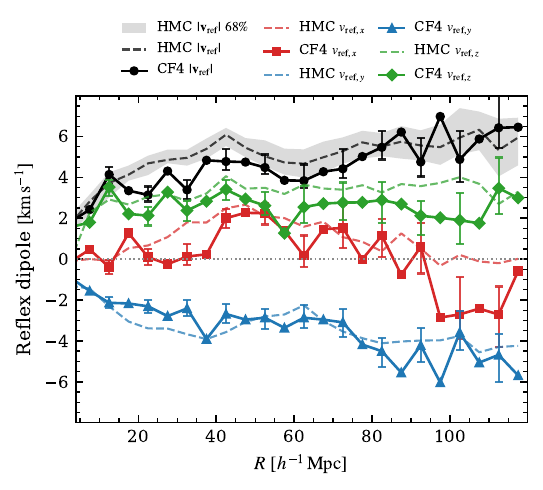}
      \caption{CF4 shell-by-shell reflex dipole compared with HMC predictions using the same tracer positions, shell selection, and dipole-fitting procedure. Panels show the amplitude and Cartesian supergalactic components. CF4 error bars come from Monte Carlo perturbations of the measured radial velocities. Dashed curves show the HMC median, and the grey band gives the central 68\% HMC interval for the amplitude.}
    \label{fig:reflex-dipole-shells}
\end{figure}

Because the CF4 sampling is non-uniform on the sky and with depth
(Section~\ref{sec:data-cf4}), we do not compare the CF4 reflex dipole
with an ideal full-sky shell average. Instead, for each HMC realization, the
Zel'dovich velocity field is interpolated to the same CF4 tracer
positions in redshift space. The same radial-shell selection is then
imposed and the same line-of-sight dipole fit is repeated.
\Cref{fig:reflex-dipole-shells} compares the resulting shell-by-shell
CF4 estimates with the corresponding HMC predictions. The total CF4
reflex-dipole amplitude follows the HMC median closely and rises to a
value comparable to the CMB dipole amplitude at the largest radii. The
component decomposition is in Cartesian supergalactic coordinates. The
${\rm SGY}$ component, which dominates the large-scale direction, agrees
particularly well between CF4 and HMC. The ${\rm SGX}$ component is the
smallest: it is consistent with zero for CF4 inside $\sim40\hmpc$, rises
with radius in both CF4 and HMC, and then turns over near $\sim90\hmpc$.
At larger distances the HMC median declines toward zero, whereas the CF4
estimate becomes negative at roughly the $-200\kms$ level, although with
large error bars. The large-radius CF4 vector is approximately
$(-200,-500,250)\kms$ in $(\mathrm{SGX},\mathrm{SGY},\mathrm{SGZ})$.
This differs from the asymptotic reflex velocity expected from the Local
Sheet motion, $-\mathbf{V}_\textrm{LS}$ in
\cref{eq:local-sheet-cmb-velocity}. In particular, both are dominated by
a negative-SGY component, but the SGX component has the opposite sign.
This mismatch is not surprising: convergence of the clustering or
velocity dipole to the CMB dipole is expected only at substantially
larger depths, of order $\gssim250\hmpc$, and depends on the survey
window and depth \citep{Nusser2014,bilicki11}.

\section{Discussion}
\label{sec:discussion}

We have presented a constrained reconstruction of the nearby Universe
from 2MRS using a Zel'dovich forward model, an explicit redshift-space
Poisson likelihood, and a nonlinear radius-dependent bias prescription.
The MDPL2 tests show that the method recovers halo radial velocities with
small  bias relative to the conditional scatter. The observational
CF4 comparisons then provide three complementary tests of the same
reconstructed velocity field: the object-by-object conditional mean, the
density--radial-velocity correlation, and the reflex dipole. Taken
together, these tests support the gravitational-instability picture in
which the large-scale velocity field is generated by the mass distribution
traced by 2MRS.

The MDPL2 tests also compare the MAP solutions obtained with the
Poisson point-process likelihood and with the Gaussian grid-based
likelihood. For the statistics shown in Fig.~\ref{fig:mdpl2-zeld-n128-fr1-fr005}, the
two MAP reconstructions differ only mildly, although the Gaussian case
shows slightly larger scatter. This comparison should not be interpreted
as a full posterior-level comparison of the two likelihoods. We have not
run and compared full HMC chains for both likelihood choices.

The object-by-object comparison should therefore be read as a test of the
mean CF4 velocity at fixed MAP prediction, as shown by the binned relation
in \Cref{fig:v-v}. The unresolved small-scale motions, distance-indicator
errors, and residual nonlinear velocities appear as scatter about this
conditional mean rather than as a failure of the reconstruction.

The residual-velocity correlation is not expected to vanish
perfectly. CF4 combines heterogeneous distance indicators, has a
nonuniform sky and depth selection, and includes residual nonlinear
motions. The model is also deliberately approximate: it uses ZA rather
than a full nonlinear solver, an effective galaxy-bias prescription, and
a simplified but explicit treatment of the survey selection and mask.
It would be useful to pinpoint the specific causes of the residual
correlations, especially on scales $\lesssim 10\hmpc$, where nonlinear
motions, shell crossing, distance errors, and limitations of the ZA model
are all expected to contribute. Nevertheless, given the complexity of the
comparison, these residuals are of secondary importance relative to the
overall agreement in the mean relation, the density--velocity correlation,
and the reflex-dipole comparison.

An important limitation of this paper is that all calculations are
restricted to a fixed, standard $\Lambda$CDM cosmology. We do not perform
cosmological parameter estimation. A full field-level inference \citep[e.g.,][]{Schmidt2020FieldLevelLikelihood,Nguyen2020EFTLikelihood,Boruah2024KARMMA} of
cosmological parameters would require sampling both the parameters and the
latent field, and then marginalizing over the sampled field (in our
parameterization, the whitened initial density).
This is a substantially
larger calculation than finding and testing a fiducial MAP reconstruction,
and is beyond the scope of the present work.

For the same reason, it would be incorrect to infer cosmological
parameters by simply rescaling the MAP velocities. The velocity associated
with a fixed redshift-space galaxy distribution does not scale linearly
with the growth rate $f(\Omega)$ in the same way as the velocity inferred
from a fixed real-space density field. Changing $f(\Omega)$ within
$\Lambda$CDM also changes the assumed power spectrum and therefore the
prior over initial conditions. Only in the restricted linear-bias and linear theory 
the relevant dependence is via the  combination
$f(\Omega)/b$, so that a change in $f(\Omega)$ at fixed cosmology can be
reinterpreted as a change in the linear bias factor $b$. That limit is not
the regime of interest in the current paper. The reconstruction uses small and mildly
nonlinear scales where a linear bias prescription is inaccurate and can
even imply unphysical negative galaxy densities.

A related modeling choice is that we do not collapse observed Fingers of
God before applying the reconstruction. Collapsing groups and clusters in
redshift space would require identifying extended radial structures and
replacing galaxies spanning a large radial and angular extent by a single
point. For the present 2MRS application this seemed too aggressive and
risked erasing real angular information. Taking Virgo as an example, the angular extent is $\approx 20^\circ$ and the radial extent corresponds to $\approx 1300\kms$.
Leaving the Fingers of God
uncollapsed preserves the angular structure of the data, at the cost of
producing somewhat fragmented cluster regions in the redshift-space
comparison. The subsequent $N$-body evolution of posterior samples shows
that nonlinear Fingers of God are restored naturally in redshift space,
which provides a useful consistency check on this choice.
A more detailed comparison focused on the Coma, Virgo, and Local Group
regions is beyond the scope of the paper. Nevertheless, halos identified by a Friends-of-Friends algorithm in the
 constrained simulations  include several large systems at the locations of the main clusters, with total masses
comparable to observational estimates for the corresponding systems, and
the projected redshift-space distributions match the 2MRS observations of
these cluster regions. The match to the Local Group neighborhood however is more subtle. 
Within a few $h^{-1}{\rm Mpc}$ of the observer, the mean density in the
realizations can be larger by a factor of roughly two to three than
nearby-galaxy estimates of the local density, although the Local Void is
still present \citep{Karachentsev2013,Tully2008,Tully2009,PeeblesTully2013}.
One possible origin of this excess is the  simple tracer
model adopted here. All 2MRS galaxies enter the likelihood on equal
footing, independent of luminosity or stellar mass, whereas the immediate
Local Group environment is dominated by low-mass galaxies. This may
therefore overweight nearby small galaxies as mass tracers and exaggerate
the inferred local overdensity.

The present analysis uses only the 2MRS redshift survey in the
reconstruction itself. 
In contrast, peculiar-velocity catalogs, like CF4, have already
been used in constrained Local-Universe simulations and Local Group
analyses. The CLUES, HESTIA, and related programs use peculiar-velocity
constraints to reconstruct the nearby large-scale structure and to build
Milky Way--M31 analogues
\citep{GottloeberHoffmanYepes2010,CarlesiEtAl2016,hestia,Valade2024Basins}. More focused
Local Group studies, including the Bayesian analysis of \citet{Wempe2024},
use local kinematic constraints on the Milky Way--M31 system and the
quiet local Hubble flow.

In the present paper CF4 is used only as an external validation data set,
not as a constraint in the likelihood. Incorporating it directly would
require a careful treatment of distance-indicator errors and
inhomogeneous Malmquist bias, especially because peculiar-velocity
uncertainties grow with distance and the catalogue cannot be treated as a
redshift survey. For this reason, 2MRS is expected to provide most of the
constraints over the larger volume, while CF4 should have its strongest
impact on the nearby velocity field and on the smallest scales where
accurate distances are available for a substantial fraction of galaxies.
In future work we intend to combine the complementary information in 2MRS
and CF4 in a joint likelihood.

A more complete statistical characterization of the reconstructed
fields is left for future work. We have not shown here the power spectra
or one-point probability density functions of the initial and evolved
fields, since a detailed analysis of these quantities would move the
paper away from its main goal of testing the 2MRS velocity prediction
with CF4. In internal checks, the power
spectrum of the HMC realizations follows the theoretical input prior well
over the resolved scales, while the MAP power is suppressed as expected
for a conditional optimum. The one-point PDF of the initial field, or
more directly of the whitened variable $\epsilon({\bf x})$, is close to
Gaussian. Since HMC samples the posterior rather than the pure Gaussian
prior, some deviations from an exactly Gaussian one-point PDF are expected
where the data constrain the field. A fuller presentation of these
power-spectrum and PDF diagnostics will be given elsewhere.


\section{Acknowledgments}
\label{sec:acknowledgments}

Special thanks to Brent Tully for many useful discussions.
This research was supported by a grant (\#893/22) from the Israel
Science Foundation and by a grant from the Asher Space Research
Institute. Additional support came from the Munich Institute for
Astro-, Particle and BioPhysics (MIAPbP), which is funded by the
Deutsche Forschungsgemeinschaft (DFG, German Research Foundation)
under Germany's Excellence Strategy -- EXC-2094 -- 390783311.


\bibliography{references,references_chat}{}

@article{SBr99,
    title = {{A First Comparison of the Surface Brightness Fluctuation Survey Distances with the Galaxy Density Field: Implications for H{\_}{\{}0{\}} and {\{}{$\Omega$}{\}}}},
    year = {1999},
    journal = {{\textbackslash}apjl},
    author = {Blakeslee, J.~P. and Davis, M and Tonry, J.~L. and Dressler, A and Ajhar, E.~A.},
    month = {12},
    pages = {L73-L76},
    volume = {527},
    doi = {10.1086/312404}
}

@article{Frisch,
    title = {{A reconstruction of the initial conditions of the Universe by optimal mass transportation}},
    year = {2002},
    journal = {\nat},
    author = {Frisch, U and Matarrese, S and Mohayaee, R and Sobolevski, A},
    month = {5},
    pages = {260--262},
    volume = {417}
}

@article{Nusser2020,
    title = {{Biasing Relation, Environmental Dependencies, and Estimation of the Growth Rate from Star-forming Galaxies}},
    year = {2020},
    journal = {ApJ},
    author = {Nusser, Adi and Yepes, Gustavo and Branchini, Enzo},
    number = {1},
    month = {12},
    pages = {47},
    volume = {905},
    publisher = {American Astronomical Society},
    url = {https://ui.adsabs.harvard.edu/abs/2020ApJ...905...47N/abstract},
    doi = {10.3847/1538-4357/abc42f},
    issn = {0004-637X}
}

@article{NBDL,
    title = {{Bulk Flows from Galaxy Luminosities: Application to 2Mass Redshift Survey and Forecast for Next-generation Data Sets}},
    year = {2011},
    journal = {ApJ},
    author = {Nusser, A and Branchini, E and Davis, M},
    month = {7},
    pages = {77},
    volume = {735},
    doi = {10.1088/0004-637X/735/2/77},
    arxivId = {astro-ph.CO/1102.4189}
}

@article{Saulder_Fundamental,
    title = {{Calibrating the fundamental plane with SDSS DR8 data}},
    year = {2013},
    journal = {AA},
    author = {Saulder, Christoph and Mieske, Steffen and Zeilinger, Werner W. and Chilingarian, Igor},
    pages = {A21},
    volume = {557},
    url = {https://ui.adsabs.harvard.edu/abs/2013A&A...557A..21S/abstract},
    doi = {10.1051/0004-6361/201321466},
    issn = {00046361},
    arxivId = {1306.0285}
}

@article{k87,
    title = {{Clustering in real space and in redshift space}},
    year = {1987},
    journal = {MNRAS},
    author = {Kaiser, N},
    month = {7},
    pages = {1--21},
    volume = {227}
}

@article{LilowNusser2021,
    title = {{Constrained realizations of 2MRS density and peculiar velocity fields: growth rate and local flow}},
    year = {2021},
    journal = {MNRAS},
    author = {Lilow, Robert and Nusser, Adi},
    number = {2},
    month = {8},
    pages = {1557--1581},
    volume = {507},
    publisher = {Oxford University Press (OUP)},
    url = {https://ui.adsabs.harvard.edu/abs/2021MNRAS.507.1557L/abstract},
    doi = {10.1093/MNRAS/STAB2009},
    issn = {0035-8711},
    arxivId = {arXiv:2102.07291}
}

@article{hr91,
    title = {{Constrained realizations of Gaussian fields - A simple algorithm}},
    year = {1991},
    journal = {ApJL},
    author = {Hoffman, Y and Ribak, E},
    month = {10},
    pages = {L5-L8},
    volume = {380},
    doi = {10.1086/186160}
}

@article{l10,
    title = {{Cosmic Flow From Two Micron All-Sky Redshift Survey: the Origin of Cosmic Microwave Background Dipole and Implications for {\{}{$\Lambda$}{\}}CDM Cosmology}},
    year = {2010},
    journal = {ApJ},
    author = {Lavaux, G and Tully, R.~B. and Mohayaee, R and Colombi, S},
    month = {1},
    pages = {483--498},
    volume = {709},
    doi = {10.1088/0004-637X/709/1/483},
    arxivId = {0810.3658}
}

@article{Lind05,
    title = {{Cosmic growth history and expansion history}},
    year = {2005},
    journal = {PRD},
    author = {Linder, Eric V.},
    number = {4},
    month = {8},
    pages = {043529},
    volume = {72},
    url = {http://link.aps.org/doi/10.1103/PhysRevD.72.043529},
    doi = {10.1103/PhysRevD.72.043529},
    issn = {1550-7998}
}

@article{CF4,
    title = {{Cosmicflows-4}},
    year = {2023},
    journal = {ApJ},
    author = {Tully, R. Brent and Kourkchi, Ehsan and Courtois, Hélène M. and Anand, Gagandeep S. and Blakeslee, John P. and Brout, Dillon and Jaeger, Thomas de and Dupuy, Alexandra and Guinet, Daniel and Howlett, Cullan and Jensen, Joseph B. and Pomar{\`{e}}de, Daniel and Rizzi, Luca and Rubin, David and Said, Khaled and Scolnic, Daniel and Stahl, Benjamin E.},
    number = {1},
    month = {2},
    pages = {94},
    volume = {944},
    publisher = {American Astronomical Society},
    url = {https://ui.adsabs.harvard.edu/abs/2023ApJ...944...94T/abstract},
    doi = {10.3847/1538-4357/ac94d8},
    issn = {0004-637X},
    arxivId = {2209.11238}
}

@article{Kourkchi2020,
    title = {{Cosmicflows-4: The Catalog of 10,000 Tully-Fisher Distances}},
    year = {2020},
    journal = {ApJ},
    author = {Kourkchi, Ehsan and Tully, R. Brent and Eftekharzadeh, Sarah and Llop, Jordan and Courtois, Hélène M. and Guinet, Daniel and Dupuy, Alexandra and Neill, James D. and Seibert, Mark and Andrews, Michael and Chuang, Juana and Danesh, Arash and Gonzalez, Randy and Holthaus, Alexandria and Mokelke, Amber and Schoen, Devin and Urasaki, Chase and Kourkchi, Ehsan and Tully, R. Brent and Eftekharzadeh, Sarah and Llop, Jordan and Courtois, Hélène M. and Guinet, Daniel and Dupuy, Alexandra and Neill, James D. and Seibert, Mark and Andrews, Michael and Chuang, Juana and Danesh, Arash and Gonzalez, Randy and Holthaus, Alexandria and Mokelke, Amber and Schoen, Devin and Urasaki, Chase},
    number = {2},
    month = {10},
    pages = {145},
    volume = {902},
    publisher = {American Astronomical Society},
    url = {https://ui.adsabs.harvard.edu/abs/2020ApJ...902..145K/abstract},
    doi = {10.3847/1538-4357/ABB66B},
    issn = {0004-637X},
    arxivId = {arXiv:2009.00733}
}

@article{carrick15,
    title = {{Cosmological parameters from the comparison of peculiar velocities with predictions from the 2M++ density field}},
    year = {2015},
    journal = {MNRAS},
    author = {Carrick, Jonathan and Turnbull, Stephen J. and Lavaux, Guilhem and Hudson, Michael J.},
    number = {1},
    month = {4},
    pages = {317--332},
    volume = {450},
    url = {http://arxiv.org/abs/1504.04627 http://dx.doi.org/10.1093/mnras/stv547},
    doi = {10.1093/mnras/stv547},
    issn = {13652966},
    arxivId = {1504.04627}
}

@article{Fundamental_Plan87,
    title = {{Fundamental Properties of Elliptical Galaxies}},
    year = {1987},
    journal = {ApJ},
    author = {Djorgovski, S. and Davis, Marc},
    month = {2},
    pages = {59},
    volume = {313},
    publisher = {American Astronomical Society},
    url = {https://ui.adsabs.harvard.edu/abs/1987ApJ...313...59D/abstract},
    doi = {10.1086/164948},
    issn = {0004-637X}
}

@article{Zehavi2011,
    title = {{GALAXY CLUSTERING IN THE COMPLETED SDSS REDSHIFT SURVEY: THE DEPENDENCE ON COLOR AND LUMINOSITY}},
    year = {2011},
    journal = {\apj},
    author = {Zehavi, Idit and Zheng, Zheng and Weinberg, David H. and Blanton, Michael R. and Bahcall, Neta A. and Berlind, Andreas A. and Brinkmann, Jon and Frieman, Joshua A. and Gunn, James E. and Lupton, Robert H. and Nichol, Robert C. and Percival, Will J. and Schneider, Donald P. and Skibba, Ramin A. and Strauss, Michael A. and Tegmark, Max and York, Donald G.},
    number = {1},
    month = {7},
    pages = {59},
    volume = {736},
    url = {http://adsabs.harvard.edu/abs/2011ApJ...736...59Z},
    doi = {10.1088/0004-637X/736/1/59},
    issn = {0004-637X}
}

@article{zeld70,
    title = {{Gravitational instability: An approximate theory for large density perturbations.}},
    year = {1970},
    journal = {A{\&}A},
    author = {Zel'dovich, Ya. B.},
    pages = {84},
    volume = {5},
    url = {http://adsabs.harvard.edu/abs/1970A%26A.....5...84Z}
}

@article{puebla16,
    title = {{Halo and subhalo demographics with Planck cosmological parameters: Bolshoi-Planck and MultiDark-Planck simulations}},
    year = {2016},
    journal = {MNRAS},
    author = {Rodr{\'{i}}guez-Puebla, Aldo and Behroozi, Peter and Primack, Joel and Klypin, Anatoly and Lee, Christoph and Hellinger, Doug},
    number = {1},
    month = {10},
    pages = {893--916},
    volume = {462},
    publisher = {Oxford University Press},
    url = {https://ui.adsabs.harvard.edu/abs/2016MNRAS.462..893R/abstract},
    doi = {10.1093/mnras/stw1705},
    issn = {13652966},
    arxivId = {1602.04813}
}

@article{bilicki11,
    title = {{Is the Two Micron All Sky Survey Clustering Dipole Convergent?}},
    year = {2011},
    journal = {ApJ},
    author = {Bilicki, M and Chodorowski, M and Jarrett, T and Mamon, G.~A.},
    month = {11},
    pages = {31},
    volume = {741},
    doi = {10.1088/0004-637X/741/1/31},
    arxivId = {astro-ph.CO/1102.4356}
}

@article{Buchert1993,
    title = {{Lagrangian theory of gravitational instability of Friedman-Lemaitre cosmologies -- second-order approach: an improved model for non-linear clustering}},
    year = {1993},
    journal = {MNRAS},
    author = {Buchert, Thomas and Ehlers, Jurgen},
    pages = {375},
    volume = {264},
    url = {http://adsabs.harvard.edu/abs/1993MNRAS.264..375B}
}

@article{Punya2023,
    title = {{Large-scale density and velocity field reconstructions with neural networks}},
    year = {2023},
    journal = {MNRAS},
    author = {Veena, Punyakoti Ganeshaiah and Lilow, Robert and Nusser, Adi},
    number = {4},
    month = {7},
    pages = {5291--5307},
    volume = {522},
    publisher = {Oxford University Press},
    url = {https://ui.adsabs.harvard.edu/abs/2023MNRAS.522.5291G/abstract},
    doi = {10.1093/mnras/stad1222},
    issn = {13652966},
    arxivId = {2212.06439}
}

@article{Klypin2016,
    title = {{Multidark simulations: The story of dark matter halo concentrations and density profiles}},
    year = {2016},
    journal = {MNRAS},
    author = {Klypin, Anatoly and Yepes, Gustavo and Gottl{\"{o}}ber, Stefan and Prada, Francisco and He{\ss}, Steffen},
    number = {4},
    month = {4},
    pages = {4340--4359},
    volume = {457},
    url = {https://academic.oup.com/mnras/article-lookup/doi/10.1093/mnras/stw248},
    doi = {10.1093/mnras/stw248},
    issn = {13652966},
    arxivId = {1411.4001}
}

@article{Lilow2024,
    title = {{Neural network reconstruction of density and velocity fields from the 2MASS Redshift Survey}},
    year = {2024},
    journal = {A{\&}A},
    author = {Lilow, Robert and Ganeshaiah Veena, Punyakoti and Nusser, Adi},
    pages = {A226},
    volume = {689},
    url = {https://ui.adsabs.harvard.edu/abs/2024arXiv240402278L/abstract},
    doi = {10.1051/0004-6361/202450219},
    arxivId = {arXiv:2404.02278}
}

@article{Riess1998,
    title = {{Observational Evidence from Supernovae for an Accelerating Universe and a Cosmological Constant}},
    year = {1998},
    journal = {AJ},
    author = {Riess, Adam G. and Filippenko, Alexei V. and Challis, Peter and Clocchiatti, Alejandro and Diercks, Alan and Garnavich, Peter M. and Gilliland, Ron L. and Hogan, Craig J. and Jha, Saurabh and Kirshner, Robert P. and Leibundgut, B. and Phillips, M. M. and Reiss, David and Schmidt, Brian P. and Schommer, Robert A. and Smith, R. Chris and Spyromilio, J. and Stubbs, Christopher and Suntzeff, Nicholas B. and Tonry, John},
    number = {3},
    month = {9},
    pages = {1009--1038},
    volume = {116},
    publisher = {American Astronomical Society},
    url = {https://ui.adsabs.harvard.edu/abs/1998AJ....116.1009R/abstract},
    doi = {10.1086/300499},
    issn = {00046256},
    arxivId = {astro-ph/9805201}
}

@article{Nusser2000,
    title = {{On the least action principle in cosmology}},
    year = {2000},
    journal = {MNRAS},
    author = {Nusser, Adi and Branchini, Enzo},
    number = {3},
    pages = {587--595},
    volume = {313},
    url = {http://adsabs.harvard.edu/cgi-bin/nph-data_query?bibcode=2000MNRAS.313..587N&link_type=ABSTRACT\npapers2://publication/doi/10.1046/j.1365-8711.2000.03261.x},
    doi = {10.1046/j.1365-8711.2000.03261.x},
    issn = {00358711}
}

@article{ND94,
    title = {{On the prediction of velocity fields from redshift space galaxy samples}},
    year = {1994},
    journal = {ApJL},
    author = {Nusser, A and Davis, M},
    month = {1},
    pages = {L1-L4},
    volume = {421},
    doi = {10.1086/187172}
}

@article{Nusser2014,
    title = {{ON THE RECOVERY OF THE LOCAL GROUP MOTION FROM GALAXY REDSHIFT SURVEYS}},
    year = {2014},
    journal = {ApJ},
    author = {Nusser, Adi and Davis, Marc and Branchini, Enzo},
    number = {2},
    month = {6},
    pages = {157},
    volume = {788},
    url = {http://adsabs.harvard.edu/abs/2014ApJ...788..157N},
    doi = {10.1088/0004-637X/788/2/157},
    issn = {0004-637X}
}

@article{BEN02,
    title = {{Peculiar velocity reconstruction with the fast action method: tests on mock redshift surveys}},
    year = {2002},
    journal = {MNRAS},
    author = {Branchini, E and Eldar, A and Nusser, A},
    month = {9},
    pages = {53--72},
    volume = {335},
    doi = {10.1046/j.1365-8711.2002.05611.x}
}

@article{KN16,
    title = {{Performance study of Lagrangian methods: reconstruction of large scale peculiar velocities and baryonic acoustic oscillations}},
    year = {2017},
    journal = {MNRAS},
    author = {Keselman, Ariel and Nusser, Adi},
    pages = {1915--1928},
    volume = {467},
    url = {https://oup.silverchair-cdn.com/oup/backfile/Content_public/Journal/mnras/467/2/10.1093_mnras_stx152/3/stx152.pdf?Expires=1496916402&Signature=be9k9p1BOxH11cgHfkkg~icueVcmsriLpBwBUnGi2QK93k~lyiYDcFqsD8aOi9~NUva9UC9-N5Zrp~h0b9k2FCOZ0Gauv61pniPgQpZfAsyorZjz},
    doi = {10.1093/mnras/stx152},
    issn = {0035-8711},
    arxivId = {1609.03576}
}

@article{Norberg2001,
    title = {{The 2dF Galaxy Redshift Survey: luminosity dependence of galaxy clustering}},
    year = {2001},
    journal = {MNRAS},
    author = {Norberg, Peder and Baugh, Carlton M. and Hawkins, Ed and Maddox, Steve and Peacock, John A. and Cole, Shaun and Frenk, Carlos S. and Bland-Hawthorn, Joss and Bridges, Terry and Cannon, Russell and Colless, Matthew and Collins, Chris and Couch, Warrick and Dalton, Gavin and De Propris, Roberto and Driver, Simon P. and Efstathiou, George and Ellis, Richard S. and Glazebrook, Karl and Jackson, Carole and Lahav, Ofer and Lewis, Ian and Lumsden, Stuart and Madgwick, Darren and Peterson, Bruce A. and Sutherland, Will and Taylor, Keith},
    number = {1},
    month = {11},
    pages = {64--70},
    volume = {328},
    url = {http://adsabs.harvard.edu/abs/2001MNRAS.328...64N},
    doi = {10.1046/j.1365-8711.2001.04839.x},
    issn = {00358711}
}

@article{2mrs2012,
    title = {{The 2MASS Redshift Survey{\{}{\textbackslash}mdash{\}}Description and Data Release}},
    year = {2012},
    journal = {ApJS},
    author = {Huchra, J.~P. and Macri, L.~M. and Masters, K.~L. and Jarrett, T.~H. and Berlind, P and Calkins, M and Crook, A.~C. and Cutri, R and Erdogdu, P and Falco, E and George, T and Hutcheson, C.~M. and Lahav, O and Mader, J and Mink, J.~D. and Martimbeau, N and Schneider, S and Skrutskie, M and Tokarz, S and Westover, M},
    month = {4},
    pages = {26},
    volume = {199},
    doi = {10.1088/0067-0049/199/2/26},
    arxivId = {astro-ph.CO/1108.0669}
}

@article{hestia,
    title = {{The hestia project: Simulations of the Local Group}},
    year = {2020},
    journal = {MNRAS},
    author = {Libeskind, Noam I. and Carlesi, Edoardo and Grand, Robert J.J. and Khalatyan, Arman and Knebe, Alexander and Pakmor, Ruediger and Pilipenko, Sergey and Pawlowski, Marcel S. and Sparre, Martin and Tempel, Elmo and Wang, Peng and Courtois, Hélène M. and Gottl{\"{o}}ber, Stefan and Hoffman, Yehuda and Minchev, Ivan and Pfrommer, Christoph and Sorce, Jenny G. and Springel, Volker and Steinmetz, Matthias and Tully, R. Brent and Vogelsberger, Mark and Yepes, Gustavo},
    number = {2},
    month = {10},
    pages = {2968--2983},
    volume = {498},
    publisher = {Oxford University Press},
    url = {https://ui.adsabs.harvard.edu/abs/2020MNRAS.498.2968L/abstract},
    doi = {10.1093/mnras/staa2541},
    issn = {13652966},
    arxivId = {2008.04926}
}

@book{Peeb80,
    title = {{The large-scale structure of the universe}},
    year = {1980},
    booktitle = {Princeton University Press},
    author = {Peebles, P. J. E.},
    publisher = {Princeton University Press, NJ},
    url = {http://adsabs.harvard.edu/abs/1980lssu.book.....P}
}

@article{BND12,
    title = {{The linear velocity field of 2MASS Redshift Survey, K<SUB>s</SUB>= 11.75 galaxies: constraints on {$\beta$} and bulk flow from the luminosity function}},
    year = {2012},
    journal = {MNRAS},
    author = {Branchini, Enzo and Davis, Marc and Nusser, Adi},
    number = {1},
    month = {7},
    pages = {472},
    volume = {424},
    publisher = {Oxford University Press},
    url = {https://ui.adsabs.harvard.edu/abs/2012MNRAS.424..472B/abstract},
    doi = {10.1111/J.1365-2966.2012.21210.X},
    issn = {0035-8711}
}

@article{a82,
    title = {{The velocity field in the local supercluster}},
    year = {1982},
    journal = {\apj},
    author = {Aaronson, M and Huchra, J and Mould, J and Schechter, P. L. and Tully, R. B.},
    month = {7},
    pages = {64},
    volume = {258},
    url = {http://adsabs.harvard.edu/doi/10.1086/160053},
    doi = {10.1086/160053},
    issn = {0004-637X}
}

@article{hht07,
    title = {{The Velocity Field of the Local Universe from Measurements of Type Ia Supernovae}},
    year = {2007},
    journal = {ApJ},
    author = {Haugb{\o}lle, T and Hannestad, S and Thomsen, B and Fynbo, J and Sollerman, J and Jha, S},
    month = {6},
    pages = {650--659},
    volume = {661},
    doi = {10.1086/513600}
}

@article{Peebles1989,
    title = {{Tracing galaxy orbits back in time}},
    year = {1989},
    journal = {ApJ},
    author = {Peebles, P. J. E.},
    month = {9},
    pages = {L53-L56},
    volume = {344},
    url = {http://adsabs.harvard.edu/doi/10.1086/185529 http://adsabs.harvard.edu/cgi-bin/nph-data_query?bibcode=1989ApJ...344L..53P&link_type=ABSTRACT\npapers2://publication/doi/10.1086/185529},
    isbn = {doi:10.1086/185529},
    doi = {10.1086/185529},
    issn = {0004-637X}
}

@article{NusserDekel92,
    title = {{Tracing large-scale fluctuations back in time}},
    year = {1992},
    journal = {\apj},
    author = {Nusser, Adi and Dekel, Avishai},
    month = {6},
    pages = {443},
    volume = {391},
    url = {http://adsabs.harvard.edu/doi/10.1086/171360},
    doi = {10.1086/171360},
    issn = {0004-637X}
}

@article{Nusser2017,
    title = {{Velocity-Density Correlations from the cosmicflows-3 Distance Catalog and the 2MASS Redshift Survey}},
    year = {2017},
    journal = {\mnras},
    author = {Nusser, Adi},
    month = {3},
    pages = {445--454},
    volume = {470},
    url = {http://arxiv.org/abs/1703.05324 http://dx.doi.org/10.1093/mnras/stx1225},
    doi = {10.1093/mnras/stx1225},
    issn = {0035-8711},
    arxivId = {1703.05324}
}

@article{Fisher95b,
    title = {{Wiener reconstruction of density, velocity and potential fields from all-sky galaxy redshift surveys}},
    year = {1995},
    journal = {MNRAS},
    author = {Fisher, K.~B. and Lahav, O and Hoffman, Y and Lynden-Bell, D and Zaroubi, S},
    month = {2},
    pages = {885--908},
    volume = {272}
}

@article{zh95,
    title = {{Wiener Reconstruction of the Large-Scale Structure}},
    year = {1995},
    journal = {ApJ},
    author = {Zaroubi, S and Hoffman, Y and Fisher, K.~B. and Lahav, O},
    month = {8},
    pages = {446-+},
    volume = {449},
    doi = {10.1086/176070}
}

@article{Mei2007,
  author = {Mei, S. and Blakeslee, J. P. and C\^ot\'e, P. and others},
  title = {The ACS Virgo Cluster Survey. XIII. SBF Distance Catalog and the Three-dimensional Structure of the Virgo Cluster},
  journal = {\apj},
  volume = {655},
  pages = {144--162},
  year = {2007},
  doi = {10.1086/509598},
  eprint = {astro-ph/0702510},
  archivePrefix = {arXiv}
}

@article{deGrijsBono2020,
  author = {de Grijs, Richard and Bono, Giuseppe},
  title = {A Distance Framework out to 100 Mpc},
  journal = {\apjs},
  volume = {251},
  number = {1},
  pages = {15},
  year = {2020},
  doi = {10.3847/1538-4365/abbb96},
  eprint = {2004.00114},
  archivePrefix = {arXiv}
}

@article{Scolnic2024,
  author = {Scolnic, Daniel and others},
  title = {The Hubble Tension in our own Backyard: DESI and the Tip of the Red Giant Branch Join the Fight over Coma},
  journal = {arXiv e-prints},
  pages = {arXiv:2409.14546},
  year = {2024},
  eprint = {2409.14546},
  archivePrefix = {arXiv}
}

@article{Wempe2024,
  author       = {Wempe, Ewoud and Lavaux, Guilhem and White, Simon D. M. and Helmi, Amina and Jasche, Jens and Stopyra, Stephen},
  title        = {Constrained cosmological simulations of the Local Group using Bayesian hierarchical field-level inference},
  journal      = {\aap},
  volume       = {691},
  pages        = {A348},
  year         = {2024},
  doi          = {10.1051/0004-6361/202450975}
}

@article{Karachentsev2013,
  author       = {Karachentsev, I. D. and Kaisina, E. I. and Makarov, D. I. and Tully, R. Brent and Rizzi, L. and Shaya, Edward J. and others},
  title        = {Updated Nearby Galaxy Catalog},
  journal      = {\aj},
  volume       = {145},
  pages        = {101},
  year         = {2013},
  doi          = {10.1088/0004-6256/145/4/101}
}

@article{Tully2009,
  author       = {Tully, R. Brent and Rizzi, Luca and Shaya, Edward J. and Courtois, H{\'e}l{\`e}ne M. and Makarov, Dmitry I. and Jacobs, Bradley A.},
  title        = {Our Peculiar Motion Away from the Local Void},
  journal      = {\aj},
  volume       = {138},
  pages        = {323--331},
  year         = {2009},
  doi          = {10.1088/0004-6256/138/2/323}
}

@article{Courtois2013,
  author       = {Courtois, H{\'e}l{\`e}ne M. and Pomar{\`e}de, Daniel and Tully, R. Brent and Hoffman, Yehuda and Courtois, Denis},
  title        = {Cosmography of the Local Universe},
  journal      = {\aj},
  volume       = {146},
  pages        = {69},
  year         = {2013},
  doi          = {10.1088/0004-6256/146/3/69}
}

@article{PeeblesTully2013,
  author       = {Peebles, P. J. E. and Tully, R. Brent},
  title        = {The Local Group as an Astrophysical Laboratory},
  journal      = {\apj},
  volume       = {778},
  pages        = {137},
  year         = {2013},
  doi          = {10.1088/0004-637X/778/2/137}
}

@ARTICLE{Tully2008,
       author = {{Tully}, R. Brent and {Shaya}, Edward J. and {Karachentsev}, Igor D. and {Courtois}, H{\'e}l{\`e}ne M. and {Kocevski}, Dale D. and {Rizzi}, Luca and {Peel}, Alan},
        title = "{Our Peculiar Motion Away from the Local Void}",
      journal = {ApJ},
         year = 2008,
        month = mar,
       volume = {676},
       number = {1},
        pages = {184-205},
          doi = {10.1086/527428},
archivePrefix = {arXiv},
       eprint = {0705.4139},
 primaryClass = {astro-ph},
       adsurl = {https://ui.adsabs.harvard.edu/abs/2008ApJ...676..184T}
}

@incollection{GottloeberHoffmanYepes2010,
  author    = {Gottl{\"o}ber, Stefan and Hoffman, Yehuda and Yepes, Gustavo},
  title     = {Constrained Local UniversE Simulations (CLUES)},
  booktitle = {High Performance Computing in Science and Engineering, Garching/Munich 2009},
  editor    = {Wagner, Siegfried and Steinmetz, Matthias and Bode, Arndt and M{\"u}ller, Markus Michael},
  publisher = {Springer Berlin Heidelberg},
  address   = {Berlin, Heidelberg},
  pages     = {309--322},
  year      = {2010},
  doi       = {10.1007/978-3-642-13872-0_26},
  eprint    = {1005.2687},
  archivePrefix = {arXiv},
  primaryClass  = {astro-ph.CO}
}

@article{CarlesiEtAl2016,
  author  = {Carlesi, Edoardo and Sorce, Jenny G. and Hoffman, Yehuda and Gottl{\"o}ber, Stefan and others},
  title   = {Constrained Local UniversE Simulations: a Local Group factory},
  journal = {MNRAS},
  volume  = {458},
  number  = {1},
  pages   = {900--911},
  year    = {2016},
  doi     = {10.1093/mnras/stw357}
}

@article{JascheWandelt2013,
  author  = {Jasche, Jens and Wandelt, Benjamin D.},
  title   = {Bayesian physical reconstruction of initial conditions from large-scale structure surveys},
  journal = {MNRAS},
  volume  = {432},
  number  = {2},
  pages   = {894--913},
  year    = {2013},
  doi     = {10.1093/mnras/stt449},
  eprint  = {1203.3639},
  archivePrefix = {arXiv},
  primaryClass  = {astro-ph.CO}
}

@article{LavauxJasche2016,
  author  = {Lavaux, Guilhem and Jasche, Jens},
  title   = {Unmasking the masked Universe: the 2M++ catalogue through Bayesian eyes},
  journal = {MNRAS},
  volume  = {455},
  number  = {3},
  pages   = {3169--3179},
  year    = {2016},
  doi     = {10.1093/mnras/stv2499},
  eprint  = {1509.05040},
  archivePrefix = {arXiv},
  primaryClass  = {astro-ph.CO}
}

@article{KitauraEtAl2012,
  author  = {Kitaura, Francisco-Shu and Erdogdu, Pirin and Nuza, Sebastian E. and Khalatyan, A. and Angulo, Raul E. and Hoffman, Yehuda and Gottl{\"o}ber, Stefan},
  title   = {Cosmic structure and dynamics of the local Universe},
  journal = {MNRAS},
  volume  = {427},
  number  = {1},
  pages   = {L35--L39},
  year    = {2012},
  doi     = {10.1111/j.1745-3933.2012.01330.x}
}

@article{Macri2019,
  author  = {Macri, Lucas M. and Kraan-Korteweg, Renee C. and Lambert, Trystan and Alonso, Maria Victoria and Berlind, Perry and Calkins, Michael and Erdogdu, Pirin and Falco, Emilio E. and Jarrett, Thomas H. and Mink, Jessica D.},
  title   = {The 2MASS Redshift Survey in the Zone of Avoidance},
  journal = {\apjs},
  volume  = {245},
  number  = {1},
  pages   = {6},
  year    = {2019},
  doi     = {10.3847/1538-4365/ab465a},
  eprint  = {1911.02944},
  archivePrefix = {arXiv},
  primaryClass  = {astro-ph.GA}
}

@article{Nikakhtar2024,
  author  = {Nikakhtar, Farnik and Sheth, Ravi K. and Padmanabhan, Nikhil and L{\'e}vy, Bruno and Mohayaee, Roya},
  title   = {Displacement field analysis via optimal transport: Multitracer approach to cosmological reconstruction},
  journal = {\prd},
  volume  = {109},
  pages   = {123512},
  year    = {2024},
  doi     = {10.1103/PhysRevD.109.123512}
}

@article{Schmidt2020FieldLevelLikelihood,
  author        = {Schmidt, Fabian},
  title         = {{An n-th order Lagrangian Forward Model for Large-Scale Structure}},
  journal       = {\jcap},
  volume        = {2021},
  number        = {04},
  pages         = {033},
  year          = {2021},
  doi           = {10.1088/1475-7516/2021/04/033},
  eprint        = {2012.09837},
  archivePrefix = {arXiv},
  primaryClass  = {astro-ph.CO}
}

@article{Nguyen2020EFTLikelihood,
  author        = {Nguyen, Nhat-Minh and Schmidt, Fabian and Lavaux, Guilhem and Jasche, Jens},
  title         = {{Impacts of the physical data model on the forward inference of initial conditions from biased tracers}},
  journal       = {\jcap},
  volume        = {2021},
  number        = {03},
  pages         = {058},
  year          = {2021},
  doi           = {10.1088/1475-7516/2021/03/058},
  eprint        = {2011.06587},
  archivePrefix = {arXiv},
  primaryClass  = {astro-ph.CO}
}

@article{Springel2021Gadget4,
  author        = {Springel, Volker and Pakmor, Ruediger and Zier, Oliver and Reinecke, Martin},
  title         = {{Simulating cosmic structure formation with the GADGET-4 code}},
  journal       = {MNRAS},
  volume        = {506},
  number        = {2},
  pages         = {2871--2949},
  year          = {2021},
  doi           = {10.1093/mnras/stab1855},
  eprint        = {2010.03567},
  archivePrefix = {arXiv},
  primaryClass  = {astro-ph.IM}
}

@article{MundowNusser2025,
  author        = {Mundow, Raeed and Nusser, Adi},
  title         = {{Estimating Cluster Masses: A Comparative Study between Machine Learning and Maximum Likelihood}},
  journal       = {\apj},
  volume        = {994},
  number        = {1},
  pages         = {38},
  year          = {2025},
  doi           = {10.3847/1538-4357/ae17c0},
  eprint        = {2507.21876},
  archivePrefix = {arXiv},
  primaryClass  = {astro-ph.CO}
}

@article{Boruah2024KARMMA,
  author        = {Boruah, Supranta S. and Fiedorowicz, Pier and Rozo, Eduardo},
  title         = {{Bayesian mass mapping with weak lensing data using KARMMA: Validation with simulations and application to Dark Energy Survey Year 3 data}},
  journal       = {\prd},
  volume        = {110},
  number        = {2},
  pages         = {023524},
  year          = {2024},
  doi           = {10.1103/PhysRevD.110.023524},
  eprint        = {2403.05484},
  archivePrefix = {arXiv},
  primaryClass  = {astro-ph.CO}
}

@article{Valade2024Basins,
  author  = {Valade, A. and Libeskind, N. I. and Pomar{\`e}de, D. and Pfeifer, S. and Hoffman, Y. and Tully, R. B. and Kourkchi, E.},
  title   = {{Identification of basins of attraction in the local Universe}},
  journal = {Nature Astronomy},
  volume  = {8},
  pages   = {1610--1616},
  year    = {2024},
  doi     = {10.1038/s41550-024-02370-0},
  eprint  = {2409.17261},
  archivePrefix = {arXiv},
  primaryClass  = {astro-ph.CO}
}

@article{McAlpine2025,
  author  = {McAlpine, S. and others},
  title   = {{Field-level inference with a Poisson galaxy likelihood}},
  journal = {\mnras},
  year    = {2025},
  doi     = {10.1093/mnras/staf767}
}

@article{Kokron2022,
  author  = {Kokron, N. and others},
  title   = {{Likelihood modelling for biased tracers in field-level inference}},
  journal = {\mnras},
  year    = {2022},
  doi     = {10.1093/mnras/stac1420}
}

@article{Nguyen2024EFT,
  author  = {Nguyen, N.-M. and others},
  title   = {{Field-level inference with effective field theory}},
  journal = {\prl},
  volume  = {133},
  pages   = {221006},
  year    = {2024},
  doi     = {10.1103/PhysRevLett.133.221006}
}

@misc{Bayer2026HEFT,
  author        = {Bayer, A. E. and others},
  title         = {{Field-level inference with hybrid effective field theory}},
  year          = {2026},
  eprint        = {2603.15732},
  archivePrefix = {arXiv},
  primaryClass  = {astro-ph.CO}
}

@misc{Akitsu2025Likelihood,
  author        = {Akitsu, K. and others},
  title         = {{Likelihood dependence in cosmological field-level inference}},
  year          = {2025},
  eprint        = {2509.09673},
  archivePrefix = {arXiv},
  primaryClass  = {astro-ph.CO}
}

@misc{Spezzati2025Likelihood,
  author        = {Spezzati, N. and others},
  title         = {{Likelihood choices and information content in field-level inference}},
  year          = {2025},
  eprint        = {2507.05378},
  archivePrefix = {arXiv},
  primaryClass  = {astro-ph.CO}
}

@inproceedings{Phan2019NumPyro,
  author    = {Phan, Du and Pradhan, Neeraj and Jankowiak, Martin},
  title     = {{Composable effects for flexible and accelerated probabilistic programming in NumPyro}},
  booktitle = {NeurIPS Program Transformations for Machine Learning Workshop},
  year      = {2019},
  eprint    = {1912.11554},
  archivePrefix = {arXiv},
  primaryClass  = {stat.ML}
}

@article{Bingham2019Pyro,
  author  = {Bingham, Eli and Chen, Jonathan P. and Jankowiak, Martin and Obermeyer, Fritz and Pradhan, Neeraj and Karaletsos, Theofanis and Singh, Rohit and Szerlip, Paul and Horsfall, Paul and Goodman, Noah D.},
  title   = {{Pyro: Deep universal probabilistic programming}},
  journal = {Journal of Machine Learning Research},
  volume  = {20},
  number  = {28},
  pages   = {1--6},
  year    = {2019}
}

@article{HoffmanGelman2014NUTS,
  author  = {Hoffman, Matthew D. and Gelman, Andrew},
  title   = {{The No-U-Turn sampler: Adaptively setting path lengths in Hamiltonian Monte Carlo}},
  journal = {Journal of Machine Learning Research},
  volume  = {15},
  number  = {47},
  pages   = {1593--1623},
  year    = {2014},
  eprint  = {1111.4246},
  archivePrefix = {arXiv},
  primaryClass  = {stat.CO}
}

@article{Desjacques2018,
  author  = {Desjacques, Vincent and Jeong, Donghui and Schmidt, Fabian},
  title   = {{Large-scale galaxy bias}},
  journal = {Physics Reports},
  volume  = {733},
  pages   = {1--193},
  year    = {2018},
  doi     = {10.1016/j.physrep.2017.12.002},
  eprint  = {1611.09787},
  archivePrefix = {arXiv},
  primaryClass  = {astro-ph.CO}
}
\bibliographystyle{aasjournal}



\end{document}